\documentclass[twocolumn]{aastex62}
\usepackage{longtable}
\usepackage{amsmath}
\usepackage{subfiles}

\graphicspath{{./}{figures/}}

\shorttitle{GRB Broad Pulse Analysis}
\shortauthors{Tak et al.}

\begin{document}

\title{Temporal and Spectral Evolution of Gamma-ray Burst Broad Pulses:\\ Identification of High Latitude Emission in the Prompt Emission.}

\author{Donggeun Tak}
\affiliation{Deutsches Elektronen-Synchrotron DESY, Platanenallee 6, 15738 Zeuthen, Germany \href{mailto:donggeun.tak@gmail.com}{donggeun.tak@gmail.com} }
\affiliation{SNU Astronomy Research Center, Seoul National University, 1 Gwanak-rho, Gwanak-gu, Seoul, Korea}
\affiliation{Korea Astronomy and Space Science Institute, Daejeon 34055, Republic of Korea}

\author{Z. Lucas Uhm}
\affiliation{Korea Astronomy and Space Science Institute, Daejeon
34055, Republic of Korea}
\author{Judith Racusin}
\affiliation{NASA Goddard Space Flight Center, Greenbelt, MD 20771, USA}
\author{Bing Zhang}
\affiliation{Nevada Center for Astrophysics, University of Nevada, Las Vegas, NV 89154, USA}
\affiliation{Department of Physics and Astronomy, University of Nevada, Las Vegas, NV 89154, USA}
\author{Sylvain Guiriec}
\affiliation{Department of Physics, The George Washington University, 725 21st Street NW, Washington, DC 20052, USA}
\affiliation{NASA Goddard Space Flight Center, Greenbelt, MD 20771, USA}
\author{Daniel Kocevski}
\affiliation{NASA Marshall Space Flight Center, Huntsville, AL 35812, USA}
\author{Bin-Bin Zhang}
\affiliation{School of Astronomy and Space Science, Nanjing University,
Nanjing 210093, China}
\author{Julie McEnery}
\affiliation{Department of Physics and Department of Astronomy, University of Maryland, College Park, MD 20742, USA}
\affiliation{NASA Goddard Space Flight Center, Greenbelt, MD 20771, USA}



\begin{abstract}
We perform a detailed analysis on broad pulses in bright Gamma-ray bursts (GRBs) to understand the evolution of GRB broad pulses. Using the temporal and spectral properties, we test the high latitude emission (HLE) scenario in the decaying phase of broad pulses. The HLE originates from the curvature effect of a relativistic spherical jet, where higher latitude photons are delayed and softer than the observer's line-of-sight emission. The signature of HLE has not yet been identified undisputedly during the prompt emission of GRBs. The HLE theory predicts a specific relation, F$_{\nu, E_{p}}$ $\propto$ E$_{p}\!^{2}$, between the peak energy $E_{p}$ in $\nu$F$_{\nu}$ spectra and the spectral flux F$_{\nu}$ measured at $E_{p}$, F$_{\nu, E_{p}}$. We search for evidence of this relation in 2157 GRBs detected by the Gamma-ray Burst Monitor (GBM) on board the {\it Fermi Gamma-ray Space Telescope} (\textit{Fermi}) from the years 2008 to 2017. After imposing unbiased selection criteria in order to minimize contamination in a signal by background and overlaps of pulses, we build a sample of 32 broad pulses in 32 GRBs. We perform a time-resolved spectral analysis on each of these 32 broad pulses and find that the evolution of 18 pulses (56\%) is clearly consistent with the HLE relation. For the 18 broad pulses, the exponent $\delta$ in the relation of F$_{\nu, E_{p}}$ $\propto$ E$_{p}\!^{\delta}$ is distributed as a Gaussian function with median and width of 1.99 and 0.34, respectively. This result provides constraint on the emission radius of GRBs with the HLE signature.
\end{abstract}


\section{Introduction} \label{sec:intro}
\begin{figure}[t]
  \centering
  \includegraphics[scale=0.70]{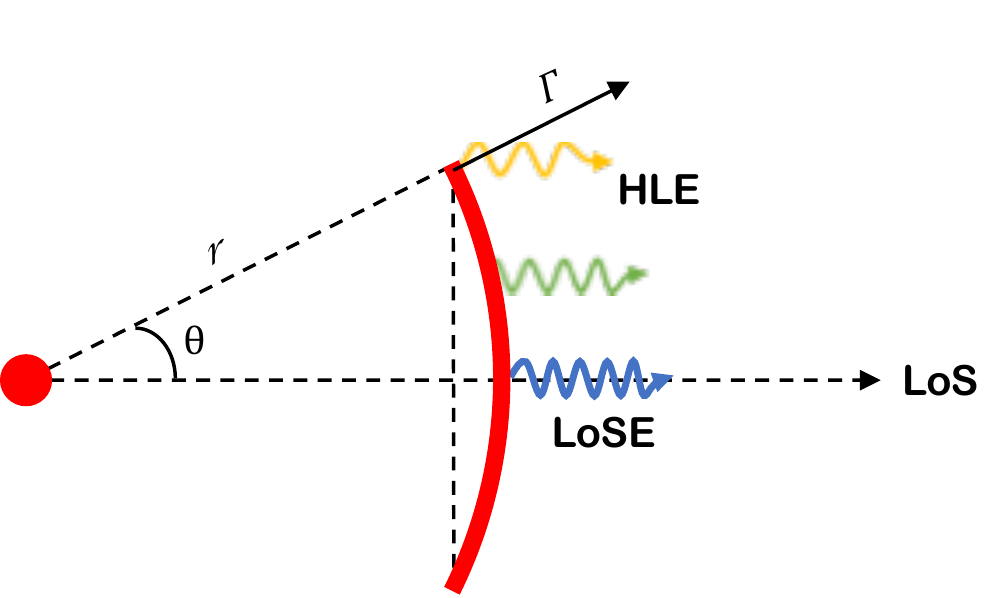}
  \caption{Schematic drawing of the geometry of an outgoing shell. LoSE and HLE stand for the line-of-sight emission and high latitude emission, respectively. If the shell expands with a Lorentz factor $\Gamma$, the photons from the LoS are boosted by 2$\Gamma$, while the photons from high latitudes are boosted by [$\Gamma$\,(1 - $\beta \cos \theta)]^{-1}$ and delayed by $\Delta$t = (r/c)\,(1- $\cos \theta)$. }
  \label{fig:geo}
\end{figure}
GRBs are the brightest electromagnetic radiation, consisting of a short gamma-ray flash (prompt emission) followed by a long-lived, broad energy band radiation (afterglow). They are attributed to emission from an outgoing relativistic jet \citep[for a recent review, see][]{Kumar2015}. Since the relativistic jet expands within a few degree \citep{Sari1999}, the geometry of the outgoing shell plays an important role in forming the observed temporal and spectral shapes (see Figure~\ref{fig:geo}). Due to the Doppler effect, emission from higher latitudes, the so-called ``high-latitude emission'' (HLE), is observed at a later time with a softer spectrum relative to the line-of-sight emission (LoSE); i.e., the higher latitude, the smaller Doppler factor. 

The relativistic curvature effect is believed to leave its signature in the 
temporal and spectral shapes of GRBs in both the prompt emission and afterglow phases \citep{Fenimore1996,Kumar2000,Dermer2004}. Through the use of numerous observations of GRB afterglows, a canonical X-ray afterglow picture has emerged \citep{Zhang2006, Nousek2006}. Among distinct afterglow phases commonly observed in the X-ray band, an early steep-decay phase connected with the tail of the prompt emission and a steep-decay phase in an X-ray flare have been interpreted as a result of the HLE effect \citep[e.g.,][]{Zhang2006,Liang2006,Yamazaki2006}. During these decaying phases, the HLE theory predicts a simple relation between the temporal index $\hat{\alpha}$ and the spectral index $\hat{\beta}$, $\hat{\alpha}$ = 2 + $\hat{\beta}$, in the convention of F$_{\nu\,{\rm obs}}$ $\propto$ t$_{\rm obs}\!^{-\hat{\alpha}}$ $\nu_{\rm obs}\!^{-\hat{\beta}}$ \citep{Kumar2000}. Note that this relation has been successfully satisfied once the onset time of emission is properly estimated \citep{Liang2006, Zhang2007, Zhang2009, Uhm2016a, Jia2016}. Also, it is suggested that an X-ray plateau emission can be HLE originated from the structured jet \citep{Oganesyan2020, Ascenzi2020}. Furthermore, the HLE evidence was also observed in the high-energy ($>$\,100 MeV) emission in GRB\,131108A \citep{Ajello2019}. 

Although such vigorous studies on HLE have been conducted, a clear HLE evidence has not been found in the prompt emission of GRBs. Observed spectral lags, the relative time difference of low-energy photons with respect to high-energy photons, have been considered as one of the HLE signatures \citep[e.g.,][]{Dermer2004, Shen2005}. However, \cite{Uhm2016b} showed that the spectral lag cannot be produced by the curvature effect but by the combination of several physical conditions such as a large emission radius, a magnetic field decreasing with radius, and a bulk acceleration. \cite{Ryde2002} and \cite{Kocevski2003b} have suggested that the asymmetric pulse shape (a fast rise and exponential decay pulse) commonly observed in the prompt phase can be interpreted as the curvature effect. They derived an analytical function for the asymmetric pulse shape and successfully fit the Burst And Transient Source Experiment (BATSE) data with the asymmetric function. \cite{Kocevski2003b} asserted that about 40\% of the BATSE sample are consistent with the predicted HLE temporal shape, although most of the others decay faster than predicted. However, as stated by the authors, this method is decoupled from the spectral property of the emission so that it cannot be irrefutable evidence.

There are several difficulties in finding the HLE signature in the prompt emission, especially in the energy spectrum. First of all, the observed temporal and spectral shapes vary one to another. Also, overlaps of multiple pulses \citep[e.g.,][]{Norris1996, Hakkila2011} and multiple spectral components in the GRB energy spectrum \citep[e.g.,][]{Abdo2009, Ackermann2010, Ackermann2011, Zhangbb2011, Guiriec2011, Guiriec2015, Tak2019} conceal the temporal and spectral properties of each pulse and the HLE signature. \cite{Genet2009} modeled the prompt emission considering the HLE effect and cautioned that the HLE relation ($\hat{\alpha}$ = 2 + $\hat{\beta}$) commonly used in the afterglow cannot be tested for a pulse in the prompt emission due to observational effects (e.g., an unclear onset time of each pulse and contamination from overlaps of nearby pulses). 

Another predicted HLE relation is a relation between the flux at a specific frequency and that frequency, $F_{\nu} (\nu_p) \propto \nu_p\!^{2}$ \citep[e.g.,][]{Dermer2004, Shenoy2013, Uhm2015}. Indeed, similar relations between the two parameters have been vigorously studied. For example, \cite{Borgonovo2001} studied the hardness-intensity correlation, $\nu F_{\nu} (E_{p}) \propto E_{p}\!^{\eta}$, with long-duration GRBs (generally $T_{90}\footnote{$T_{90}$ is a time interval containing 90\% of the background-subtracted events from a GRB} > 2 s$) detected by BATSE. Compared to the bolometric flux, the use of the flux at the peak energy, $\nu F_{\nu} (E_{p})$, can alleviate many observational limits such as limited spectral coverage of detectors and overlap of weak/soft spectral components or pulses. They found that 57\% of their sample satisfy the correlation within the prompt emission, and the exponent $\eta$ is approximately distributed as a normal distribution with median and width of 2.0 and 0.68, respectively. However, this $\eta$ value was not consistent with the HLE prediction; the predicted value for the exponent $\eta$ is 3, $\nu F_{\nu} (E_{p}) \propto E_{p}\!^{3}$. Therefore, an unambiguous conclusion has not been drawn in the prompt emission.

Recently, \cite{Li2021} tested the curvature effect with the \textit{Fermi} GRBs and showed the evidence of bulk acceleration. Uhm et al. (2022, submitted) performed a detailed numerical modeling for broad pulses in the prompt emission with a simple physical picture and provided possible scenarios of the HLE signature. In this work, we perform systematic study on a large sample of broad pulses in the \textit{Fermi}-GBM GRBs observed from 2008 to 2017 and characterize the temporal and spectral evolutional features. We also test the scaling relation, 
\begin{equation}\label{eq:hle}
    F_{\nu, E_{p}} = A E_{p}\,^{\delta}, 
\end{equation}
in order to identify the HLE signature. First of all, we select bright GRBs and impose several criteria to find relatively clean broad pulses (Section~\ref{sec:selection}). The method for identifying the HLE signature and the result are presented in Section~\ref{sec:analysis} and Section~\ref{sec:results}, respectively. In Section~\ref{sec:discussion}, we discuss several implications of the results. Finally, we conclude in Section~\ref{sec:conclusion}.

\section{Sample Selection} \label{sec:selection}
Our initial sample consists of 2157 GRBs listed in the \textit{Fermi}-GBM catalog observed in 2008--2017 \citep{Gruber2014, vonKienlin2014, Bhat2016,Yu2016}. In order to 
study the temporal and spectral features of GRB broad pulses, we perform a time-resolved analysis on a sufficient number of bins. Since only bright GRBs can provide well-constrained parameters in the time-resolved spectral analysis, we select GRBs based on energy fluence and peak flux in the energy band from 10 keV to 1 MeV; i.e., the energy fluence threshold and the peak flux threshold are 2.5 $\times$ 10\textsuperscript{-5} erg cm\textsuperscript{-2} and 1.8 $\times$ 10\textsuperscript{-6} erg cm\textsuperscript{-2} s\textsuperscript{-1}, respectively. In this selection process, we adopt the energy fluence and peak flux from the \textit{Fermi}-GBM online catalog\footnote{\href{https://heasarc.gsfc.nasa.gov/W3Browse/fermi/fermigbrst.html}{https://heasarc.gsfc.nasa.gov/W3Browse/fermi/fermigbrst.html}}, and 175 GRBs ($\sim$ 8.1 \%) survive after applying these threshold cuts. 

The \textit{Fermi}-GBM is composed of twelve NaI detectors (sodium iodide; 8 keV--1 MeV) and two BGO detectors (bismuth germanate; 200 keV--40 MeV). Among 14 GBM TTE data\footnote{Event data with a time precise to 2 microseconds in 128 energy channels.}, we configure a dataset adopting the set of detectors listed in the ``Scat Detector Mask''\footnote{A list of GBM detectors that is used in the GBM spectral catalog fits} in the GBM online catalog. For the dataset, we gather events of 50--300 keV and apply the Bayesian block algorithm, where photon-counting data are segmented based on a sudden change in a count rate \citep{Scargle2013}. We estimate a background rate for each energy channel by extrapolating the polynomial function given by fitting time intervals before and after the burst. Finally, we construct a background-subtracted count-rate curve with the Bayesian-block bins. 

We define a pulse as a series of bins where the count rates are $3\,\sigma$ above the background level. Among one or more pulses in a GRB, we focus on the brightest pulse with the highest count-rate bin; i.e., for each GRB, we select a single broad-pulse target. To define a bright, broad pulse and avoid a selection bias, we impose four criteria to systematically filter out pulses which have noticeable contaminations from background and/or overlapping pulses. Again to minimize a possible systematic bias, these criteria are not based on any physical models or analytical pulse profiles. For each criterion, we also prescribe a yellow flag (warning) or red (rule-out) flag, depending on the satisfaction level to the criterion.
\begin{enumerate}
\item The target pulse (the brightest broad pulse among pulses in the prompt emission) should contain 90\% of the GRB fluence;
\begin{itemize}
\setlength\itemsep{0.1em}
\item Yellow flag: $70\% \leq S_{p} < 90\%$,
\item Red flag: $S_{p} < 70\%$,
\end{itemize}
where $S_{p}$ is fluence of the brightest pulse in a GRB. This criterion is directly related to the primary fluence and peak flux threshold cuts. This criterion checks whether the pulse is bright enough for the time-resolved analysis. With this criterion, low-luminous GRBs composed of multiple low-fluence pulses are removed.

\item The decaying phase duration ($t_{d}$) in the target pulse should be longer than the rising phase duration ($t_{r}$);
\begin{itemize}
\item Yellow flag: $\frac{t_{r}}{2} \leq t_{d} < t_{r}$,
\item Red flag: $t_{d} < \frac{t_{r}}{2}$,
\end{itemize}
where $t_{r}$ is defined as duration from the start of the first bin of a target pulse to the peak of the target pulse, and $t_{d}$ as duration from the end of $t_{r}$ to the end of the last bin of the target pulse. 

There is no consensus in the GRB pulse shape, but a single GRB pulse is believed to be asymmetric \citep[a fast rise and exponential decay pulse (FRED);][]{Norris1996,Ryde2002, Kocevski2003b}. A pulse with $t_{d} \ll t_{r}$ is likely to be a broad pulse superposed with many FRED pulses, and this criterion is intended to exclude such pulses.

\item The decay phase of the target pulse should be clean; i.e., the count rate of bins should decrease in time without significant fluctuation. If not, the duration of bumps ($t_{b}$) should be short;
\begin{itemize}
\item Yellow flag: $t_{b} < \frac{t_{d}}{4}$,
\item Red flag: $t_{b} \geq \frac{t_{d}}{4}$, or $N_{b} \geq 2$,
\end{itemize}
where $N_{b}$ is the number of bumps. A bump is defined as a series of the irregular bins whose count rates are higher than the previous regular bin. This bump ends when a following count rate becomes lower than the bin before the start of the bump. This criterion is to minimize ambiguity due to the interplay between coexistent pulses. Any sub-dominant pulses on top of the broad pulse make it difficult to extract temporal and spectral features of the broad pulse. 

\item The target pulse should not overlap with any nearby pulses.
\begin{itemize}
\item Yellow flag: $t_{sp} \leq 10 s$,
\item Red flag: $t_{sp} \leq 5 s$,
\end{itemize}
where $t_{sp}$ is the separation time between the target pulse and nearby pulses. To check this criterion, we check the count rates of three nearby bins within 5 s and 10 s before and after the target pulse. If any nearby pulse has a count rate $3\,\sigma$ above the background level, we prescribe a yellow or red flag depending on the proximity of the pulse overlapping the target pulse.
\end{enumerate}

If the target pulse receives at least one red flag or at least two yellow flags, the pulse and corresponding GRB is removed from our sample. The results of applying these four criteria to 175 GRBs are presented in Appendix~\ref{sec:app1}. After the selection procedure, our final sample consists of 32 bright broad pulses from 32 GRBs. 

\section{Time-resolved Spectral Analysis} \label{sec:analysis}
For each broad pulse, we perform a time-resolved spectral analysis with GBM data (TTE) from the selected detectors that were employed for the sample selection (Section~\ref{sec:selection}). For each detector, low- and high-energy regimes are ignored due to overflows of channels; i.e., we use 8 keV--1 MeV for NaIs and 200 keV--40 MeV for BGOs. First of all, we estimate a background rate for each of GBM detectors with \textit{rmfit} (version43pr2)\footnote{\textit{rmfit}, \href{https://fermi.gsfc.nasa.gov/ssc/data/analysis/rmfit/}{https://fermi.gsfc.nasa.gov/ssc/data/analysis/rmfit/} for details.} by fitting a polynomial function for pre-burst and post-burst time intervals. We generate dataset by writing out the source and background regions using selected data and background fit obtained in \textit{rmfit}. Next, we perform the spectral analysis with \textit{Xspec} (version 12.10.0)\footnote{\textit{Xspec}, \href{https://heasarc.gsfc.nasa.gov/xanadu/xspec/}{https://heasarc.gsfc.nasa.gov/xanadu/xspec/} for details.}. The decaying phase of a broad pulse is divided into equally-spaced bins in  the logarithmic space (Table~\ref{tab:result}). We test three representative models for each time interval, a simple power law (PL), a power law with exponential cutoff (CPL), and the Band function \citep[Band;][]{Band1993}. The discussion on the multiple spectral component model is in Section~\ref{sec:multi}. The Poisson data with Gaussian background STATistic (PGSTAT) is adopted to estimate parameters and their errors. The best-fit model for each time interval is determined by comparing PGSTATs of each model, similar to the criteria employed by the GBM catalog \citep{Gruber2014, vonKienlin2014, Bhat2016, Yu2016}; the best-fit model is Band when $\Delta$PGSTAT (CPL - Band) $>$ 11.83 units, CPL when $\Delta$PGSTAT (PL - CPL) $>$ 8.58 units or PL otherwise.

The value and error of $F_{\nu, E_{p}}$ in CPL or Band cannot be directly computed from the parameter estimation procedure, so we compute them with the Monte Carlo simulation. We synthesize 10$^5$ spectra with parameters and corresponding covariance matrix of the best-fit model, and compute $F_{\nu, E_{p}}$ for each synthesized spectrum. From the obtained $F_{\nu, E_{p}}$ values, we obtain mean and asymmetric errors of $F_{\nu, E_{p}}$. Note that we aim to test the scaling relation between $E_{p}$ and $F_{\nu, E_{p}}$, so we ignore time intervals where the best-fit model is PL. 

Next, we perform the maximum likelihood analysis to test the scaling relation between $E_{p}$ and $F_{\nu, E_{p}}$ (Equation~\ref{eq:hle}; the bottom-left panel in Figure~\ref{fig:result}). We fit the function to all possible sets of temporally connected points, which should have at least four connected points. We use at least four points because two or three points can be aligned in a specific slope by chance but the plausibility of four points lining up at the specific value is relatively unlikely. 

To identify the HLE signature, we firstly fix the exponent to be consistent with the HLE prediction, $\delta$ = 2. For a set that is well consistent with the HLE relation ($\chi^{2}_{\nu} < 2$), we fit the same function again but leaving $\delta$ free to obtain the true exponent. When any sets of points show $\delta$ consistent with the HLE relation within the $\pm 1 \sigma$ level, we classify the bright, broad pulse as a ``clear'' case, except for cases with a large error ($\sigma_{\delta} > 1.5$). When there are multiple sets of points satisfying the relation, we take $\delta$ from the fit using the largest number of points. For a broad pulse that does not have any series of points consistent with the HLE relation, we assign the pulse to either ``weak'' or ``N/A'' after visual inspection. If a pulse shows the good agreement with the HLE prediction when excluding few outliers or shows the specific slope different from the HLE theory ($\delta \neq 2$), we mark it as a ``weak'' case. Note that one may classify these weak cases into N/A.

The HLE theory predicts not only the $E_{p}-F_{\nu, E_{p}}$ relation, but also two other scaling relations related to observer time ($t_{\rm obs}$) measured from the beginning of a broad pulse\footnote{These relations are valid only when the bulk Lorentz factor remains constant. For the case of bulk acceleration or deceleration, the exponent of such relations can change \citep{Uhm2015}.}, 
\begin{equation}\label{eq:add}
\begin{aligned}
E_{p} &\propto t_{\rm obs}\!^{-1},\\
F_{\nu, E_{p}} &\propto t_{\rm obs}\!^{-2}.\\
\end{aligned}
\end{equation}
Compared to the $E_{p}-F_{\nu, E_{p}}$ relation, it is difficult to test these relations due to the observational limitation that we do not know the true onset time of a broad pulse; the beginning of a broad pulse does not have to be same to the GRB trigger time \citep[the zero time point (t$_0$) effect;][]{Zhang2006}. To eliminate the $t_0$ effect, we define and reset the pulse start time ($t_{0, \rm obs}$) where the background-subtracted count rate starts to rise 3$\,\sigma$ above the background level (Table~\ref{tab:result}). Due to this ambiguity, these relations can only be used for double-checking and/or supporting materials of identifying the HLE evidence.

\section{Results} \label{sec:results}
\begin{figure*}[t]
  \centering
  \includegraphics[scale=0.45]{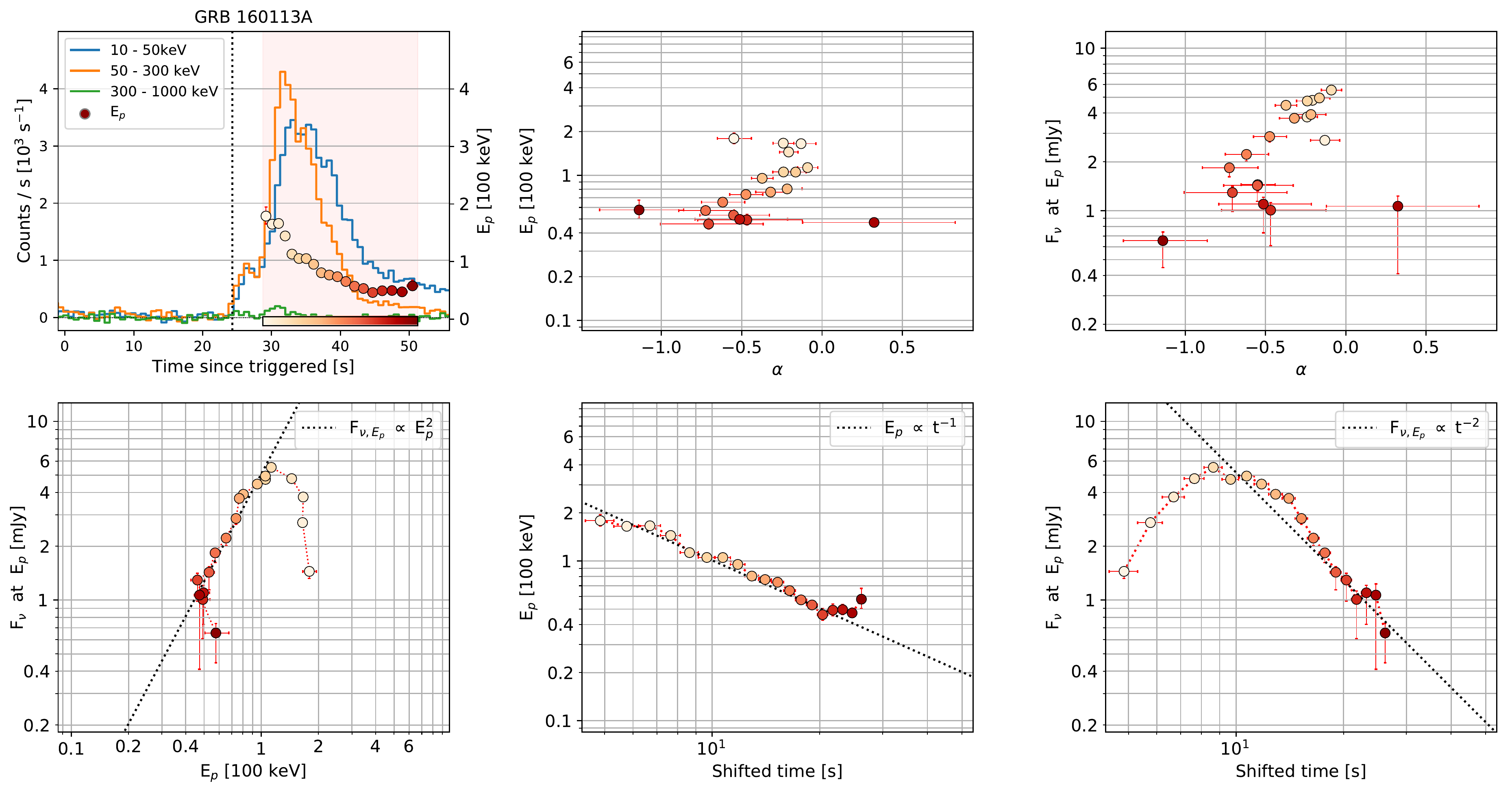}
  \caption{Spectral analysis on a broad pulse in GRB\,160113A. The top left panel shows a count rate curve with the evolution of $E_{p}$. The pink shaded region indicates a time interval where either CPL or Band is the best-fit model. The color gradation used in $E_{p}^{2}$ represents the lapse of time, and the color coding is used in the other panels. The others two top panels (upper center and upper right) show $\alpha$ vs $E_{p}$ and $\alpha$ vs $F_{\nu, E_{p}}$, respectively. The bottom panels show the agreement between data and the HLE relations: $E_{p}-F_{\nu, E_{p}}$, $t-E_{p}$ and $t-F_{\nu, E_{p}}$ from left to right. The dotted lines in the bottom panels indicate the lines predicted by the HLE theory.}
  \label{fig:result}
\end{figure*}

\begin{table*}
\small
\centering 
	\begin{tabular}{c c c c c c c c c c}
    \hline\hline
    GRB name & Pulse start & Analysis start & Analysis end & \# of bins & HLE evidence & Power-law index & $\chi^{2}_{\nu}$ \\
    & [\,s\,] & [\,s\,] & [\,s\,] & & & & \\ \hline
GRB\,081009A & -0.194 & 1.630 & 7.130 & 19 & N/A & - & - \\ 
GRB\,081221A & 16.922 & 18.290 & 35.290 & 19 & Clear & 1.8 $\pm$ 0.6 & 0.2 \\
GRB\,081224A & -0.260 & 0.500 & 28.000 & 14 & Clear & 2.0 $\pm$ 0.4 & 1.2 \\
GRB\,090717A & -0.360 & 3.400 & 16.400 & 13 & Weak & - & - \\
GRB\,090719A & -0.136 & 3.480 & 14.480 & 14 & Weak & - & - \\
GRB\,090820A & 29.494 & 32.830 & 39.830 & 21 & Clear & 2.0 $\pm$ 0.9 & 0.1 \\
GRB\,100324B & -0.066 & 1.130 & 11.630 & 21 & Clear & 2.1 $\pm$ 0.4 & 0.4 \\
GRB\,101023A & 60.842 & 60.630 & 95.130 & 15 & Clear & 1.8 $\pm$ 0.3 & 0.5 \\
GRB\,110301A & -0.086 & 1.750 & 10.250 & 19 & Clear & 1.9 $\pm$ 0.3 & 0.3 \\
GRB\,110721A & -0.132 & 0.500 & 22.500 & 14 & Clear & 2.0 $\pm$ 1.4 & 0.1 \\
GRB\,110920A & -0.076 & 5.000 & 185.500 & 29 & Clear & 2.1 $\pm$ 1.0 & 0.1 \\
GRB\,120204A & 19.636 & 29.080 & 62.580 & 19 & N/A & - & - \\
GRB\,120624B & 1.950 & 10.750 & 37.750 & 14 & Weak & - & - \\
GRB\,121122A & -0.062 & 0.170 & 11.170 & 14 & N/A & - & - \\
GRB\,130219A & 72.754 & 77.830 & 112.330 & 14 & N/A & - & - \\
GRB\,130305A & 0.498 & 1.830 & 14.830 & 9 & N/A & - & - \\
GRB\,131028A & 2.482 & 5.250 & 31.750 & 17 & Weak & - & - \\
GRB\,131214A & 56.154 & 60.250 & 81.750 & 13 & Clear & 1.7 $\pm$ 0.5 & 1.7 \\
GRB\,140206B & 5.202 & 5.830 & 27.330 & 22 & Clear & 2.3 $\pm$ 0.9 & 1.7 \\
GRB\,140329A & 19.400 & 22.500 & 28.500 & 14 & Clear & 2.2 $\pm$ 0.8 & 1.7 \\
GRB\,141028A & 6.118 & 7.630 & 39.130 & 17 & Clear & 2.4 $\pm$ 1.1 & 0.6 \\
GRB\,150213B & -0.062 & 2.170 & 5.170 & 14 & Clear & 1.9 $\pm$ 0.4 & 1.1 \\
GRB\,150306A & -0.562 & 2.750 & 12.750 & 9 & Clear & 2.0 $\pm$ 0.4 & 0.4 \\
GRB\,150403A & -0.818 & 3.250 & 29.750 & 23 & N/A & - & - \\
GRB\,150902A & -0.032 & 8.500 & 14.500 & 19 & Weak & - & - \\
GRB\,151107B & -0.050 & 7.250 & 42.750 & 12 & Clear & 2.2 $\pm$ 0.5 & 0.2 \\
GRB\,160113A & 24.334 & 28.750 & 51.250 & 19 & Clear & 2.1 $\pm$ 0.2 & 1.0 \\
GRB\,160509A & -0.626 & 8.170 & 36.670 & 16 & Clear & 2.1 $\pm$ 0.6 & 0.3 \\
GRB\,160530B & -0.536 & 3.880 & 15.880 & 19 & Clear & 2.1 $\pm$ 0.2 & 1.8 \\
GRB\,160910A & -0.078 & 6.330 & 31.830 & 19 & Weak & - & - \\
GRB\,170921B & -0.152 & 0.440 & 37.440 & 19 & N/A & - & - \\
GRB\,171210A & -0.588 & 1.500 & 108.000 & 24 & Weak & - & - \\
\hline
\end{tabular}
\caption{Spectral analysis parameters for selected GRBs. }
\label{tab:result}
\end{table*}

\begin{figure}[t]
  \centering
  \includegraphics[scale=0.6]{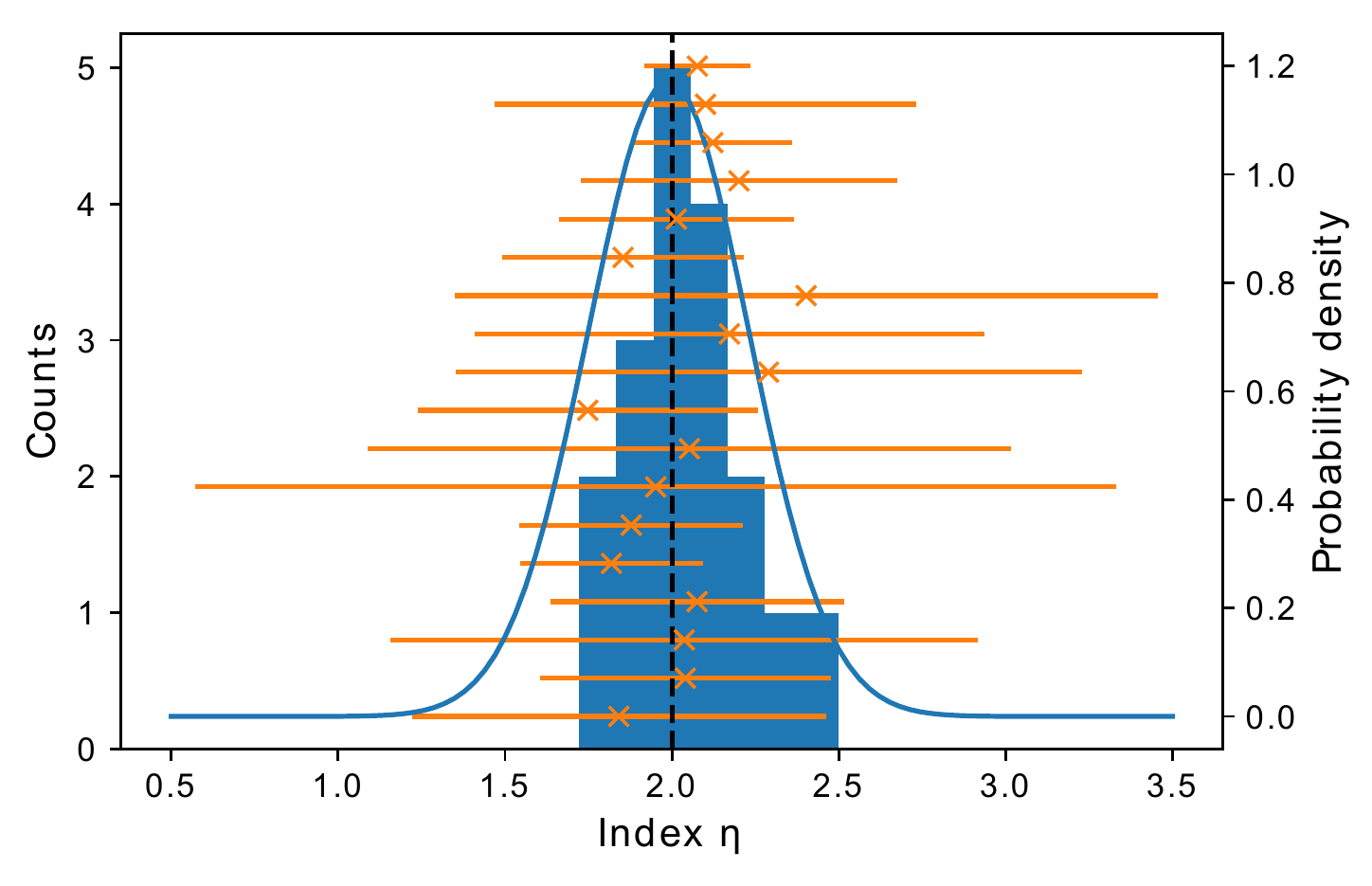}
  \caption{Distribution of $\delta$. The blue bar is the distribution of $\delta$, and the blue line shows a normal probability function given by the Monte Carlo simulation with 18 broad pulses. The median and width of the distribution is 1.99 and 0.34, respectively. The orange error bars indicate the fit results of 18 GRBs exhibiting the HLE signature clearly. The black dotted line represents the HLE expectation, $\delta$ = 2, and the data and corresponding distribution agree with the HLE theory.}
  \label{fig:dist}
\end{figure}

Figure~\ref{fig:result} shows the analysis result of GRB\,160113A as an example. The upper--left panel shows a count-rate curve in three different energy bands; 10--50 keV, 50--300 keV, and 300--1000 keV. From this panel, one can check the existence of the spectral lag and pulse shapes in different energy bands. Also, we plot the temporal evolution of $E_{p}$ on top of the count-rate curve to see the correlation between $E_{p}$ and the light curves. In most cases, either ``hard to soft'' or ``flux tracking'' pattern is observed, which is consistent with other observational studies \citep{Ford1995, Liang1996, Norris1996, Lu2012, Yu2016, Lu2018}. A low-energy photon index, $\alpha$, versus $E_{p}$ (the upper--center panel) and $\alpha$ versus $F_{\nu, E_{p}}$ (the upper--right panel) are plotted in order to see the evolution of $\alpha$ versus $E_{p}$ and F$_{\nu, E_{p}}$. The test results of three HLE relations are presented in the lower panels, $E_{p}-F_{\nu, E_{p}}$ (left), $t_{\rm obs}-E_{p}$ (center), and $t_{\rm obs}-F_{\nu, E_{p}}$ (right). The analysis results and figures for the complete sample of 32 broad pulses from 32 GRBs are in Appendix~\ref{sec:app2}.

From the $\chi^{2}$ goodness of fit test ($\chi^{2}_{\nu} < 2$ when $\delta = 2$), we find that the consistency with the HLE relation is observed in 18 out of 32 broad pulses ($\sim$ 56\%). The 18 pulses with the clear signature satisfy not only the HLE relation between $E_{p}$ and $F_{\nu, E_{p}}$, but also, in most cases, the other HLE relations, $t_{\rm obs}-E_{p}$ and $t_{\rm obs}-F_{\nu, E_{p}}$. Furthermore, as shown in Table~\ref{tab:result}, the clear cases show $\delta \sim 2$ when we let $\delta$ free. The distribution of $\delta$ clearly resembles the normal distribution (Figure~\ref{fig:dist}). We estimate median and width of the distribution from the Monte Carlo simulation: $\delta$ = $1.99\substack{+0.14 \\ -0.14}$ and $\sigma_\delta$ = 0.34 $\substack{+0.21 \\ -0.13}$, respectively. 

\section{Discussion} \label{sec:discussion}
\subsection{$t$ vs $E_{p}$ and $t$ vs $F_{\nu, E_{p}}$} \label{sec:t_relations}
The GRBs with the clear HLE signature generally satisfy all three HLE predictions. However, in few cases, we can see deviation in the scaling relations related to $t_{0, \rm obs}$ (Equation~\ref{eq:add}), especially conspicuous in $t_{obs}-F_{\nu, E_{p}}$ space; the observed data are aligned but show a slope steeper than the HLE expectation; e.g., GRB\,100324B (Figure~\ref{fig:100324B}), GRB\,110301A (Figure~\ref{fig:110301A}), and GRB\,160530B (Figure~\ref{fig:160530B}).

A simple explanation for these exceptions is because of the usage of an inaccurate value of $t_{0, \rm obs}$. The true onset time, $t_{0, \rm true}$, can not be exactly measured from observed data due to inevitable observational limits such as the background fluctuation, the detector sensitivity, and/or overlaps of pulses. The combination of these effects hides the true beginning of a broad pulse, resulting in $t_{0, \rm obs}$ later than t$_{0, \rm true}$ ($t_{0, \rm obs} \gtrsim t_{0, \rm true}$). These observational effects can lead to the deviation in the slopes of $t_{\rm obs}-E_{p}$ and $t_{\rm obs}-F_{\nu, E_{p}}$.

Another explanation for the steeper slopes can be the effect arising from the bulk acceleration \citep{Uhm2015, Uhm2016a, Uhm2018, Li2021}. 

\subsection{Interesting cases: $\delta$ of 0.7} \label{sec:otherslope}

There are two pulses showing a break in their $E_{p}-F_{\nu, E_{p}}$ curve: GRB\,110920A (clear; Figure~\ref{fig:110920A}) and GRB\,171210A (weak; Figure~\ref{fig:171210A}). For these GRBs, we compute $\delta$ for before and after the break. 

GRB\,110920A has $\delta_{110920A,\,{\rm before}} = 0.7 \pm 0.1$ and $\delta_{110920A,\, {\rm after}} = 2.1 \pm 1.0$. We emphasize that \cite{Shenoy2013} also analyzed GRB\,110920A and found that the slope $\eta$ in $(\nu F_{\nu})_{, E_{p}} \propto E_{p}\!^{\eta}$ is 1.64 $\pm$ 0.01, which is consistent with $\delta_{110920A,\, {\rm before}}$ ($\eta \simeq \delta + 1$). This implies that $\eta$ from \cite{Shenoy2013} is likely to be computed by the data points before the break. In case of GRB\,171210A, the exponents are $\delta_{171210A,\, {\rm before}} = 0.7 \pm 0.1$ and $\delta_{171210A,\, {\rm after}} = 4.0 \pm 0.3$. Interestingly, before the break, the slopes of two GRBs are consistent, $\delta \sim 0.7$. Also, we found the similar slope in GRB\,131028A (weak; Figure~\ref{fig:131028A}), $\delta_{131028A} = 0.7 \pm 0.1$. Note that GRB\,131028A is assigned to ``weak'' because the slope is not consistent with the HLE prediction. 

In these three GRBs, the common slope of $\delta \sim 0.7$ is made up with a large number of data points so that it is evident that this slope is not built up by chance. Also, this slope is observed, regardless of the existence of the HLE signature. Therefore, it is plausible that there would be another physical explanation for this slope. 

\subsection{Evolution of the low-energy photon index $\alpha$} \label{sec:alpha}
\begin{figure}[t]
  \centering
  \includegraphics[scale=0.6]{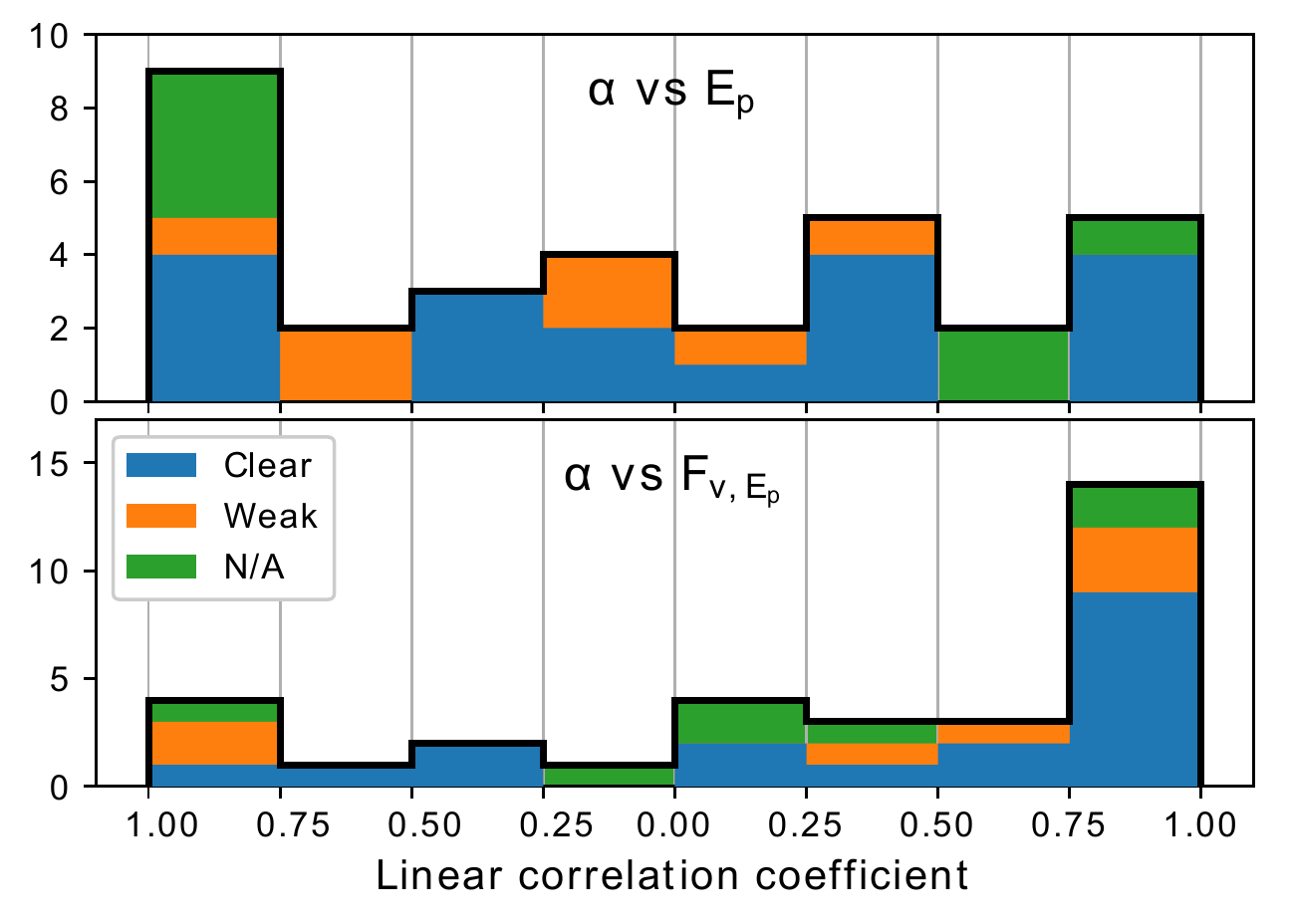}
  \caption{Stacked distributions of the linear correlation coefficients. The upper is the correlation between $\alpha$ and $E_{p}$, and the lower is $\alpha$ and F$_{\nu, E_{p}}$.}
  \label{fig:corr}
\end{figure}
We find that many GRBs show a linear correlation between parameters, $\alpha$ versus $E_{p}$ and $\alpha$ versus $F_{\nu, E_{p}}$ (See Figure~\ref{fig:result}), especially when the consistency with HLE is clear. We compute the linear correlation coefficient between such parameters for the last four time intervals. We use the last four time intervals because the HLE signature is expected to show up in late time intervals. As shown in Figure~\ref{fig:corr}, the correlation between $\alpha$ and $F_{\nu, E_{p}}$ is apparent compared to the correlation between $\alpha$ and $E_{p}$. In both the clear and weak sample, there are many cases that show the positive $\alpha-F_{\nu, E_{p}}$ correlation, whereas the N/A sample occasionally shows the positive correlation. 

This $\alpha-F_{\nu, E_{p}}$ correlation may be attributed to the curvature effect. Uhm et al. (2022, submitted) found that the peak of $\nu F_{\nu}$ spectrum is dominated by HLE, and the energy spectrum below $E_{p}$ is obtained as the combined effect of the two spectra, LoSE and HLE. This implies that $\alpha$ is not solely given by LoSE, but is affected by HLE. Therefore, the linearity of $\alpha-F_{\nu, E_{p}}$ can be related to the HLE effect. We note, though, that this conclusion is drawn from a relatively small number of data points and calls for further studies to understand the physical origin of this correlation.

\subsection{Effects of overlapping components in energy spectrum} \label{sec:multi}
In this study, we assume that each time-resolved energy spectrum is dominated by a single spectral component, which can be PL, CPL, or the Band function. We acknowledge that the best-fit for each time interval may be a combination of multiple spectral components as observed in many GRBs \citep[e.g.,][]{Guiriec2011, Guiriec2015, Tak2019}. However, if one component overwhelms the other components so that a single component model can adequately describe the peak of the energy spectrum, it is still possible to search for the HLE signature by testing of a single spectral component. The clearly observed scaling relation supports that the analysis with a single component is valid because it is difficult to explain the specific evolution of $E_{p}$ and $F_{\nu,E_{p}}$ in the presence of equally-bright multiple spectral components. With the multi-component analysis, we expect to find the HLE signature in a GRB that does not show a clear evidence in this study due to the overlaps of equally-bright multiple spectral components. However, the multi-component analysis requires a more sophisticated method to minimize systematic effects; e.g., tracking the evolution of each spectral component out of multiple components. 

\subsection{Comparison with other relations} \label{sec:otherCases}

In many observational studies, a correlation between $E_{p}$ and other physical parameters has been studied in a time-resolved spectral analysis. \cite{Borgonovo2001} performed a time-resolved analysis with GRBs detected by \textit{BATSE} and found that $(\nu F_{\nu})_{, E_{p}} \propto E_{p}\!^{2.0 \pm 0.68}$. This is not consistent with the HLE theory but may be consistent with $\delta$ $\sim$ 0.7 (equivalent to $\eta$ $\sim$ 1.7). In addition, many authors have searched for an empirical relation between the bolometric flux and $E_{p}$ \citep[e.g.,][]{Golenetskii1983, Kargatis1995, Guiriec2015}. Assuming that $(\nu F_{\nu})_{, E_{p}}$ is proportional to the bolometric flux, the slopes of many empirical relations are rather similar to $\delta = 0.7$. The reason for the discrepancy between this study (the HLE relation) and the slopes of many other empirical relations may result from the fact that the most of such empirical relations are computed around the peak of a broad pulse (the early phase), whereas our studies focus on the tail of the broad pulse. As shown in many pulses (see Appendix~\ref{sec:app2}), the exponent $\delta$ tends to be shallower in the early phase (around the peak) compared to the later phase (the tail). 

\subsection{Scaling relations from other processes} \label{sec:otherEffects}

A scaling relation between $F_{\nu} (\nu_p)$ and $\nu_p$ can result from other physical processes. \cite{Wang2000} performed a simulation study on the GRB spectrum considering the cooling of relativistic electrons in an internal shock scenario. They did not provide the scaling relation between $F_{\nu} (\nu_p)$ and $\nu_p$ but showed the temporal evolution of the peak energy in the $\nu F_\nu$ spectra \citep[Fig. 5 in][]{Wang2000}. In the figure, the energy flux at $\nu_p$ appears to decay significantly slower than that of the HLE expectation. This implies that the electron cooling alone cannot account for the observed scaling relation.

On the other hand, the synchrotron emission from the external forward shock, which is accompanied with bulk deceleration, can produce a scaling relation between $F_{\nu}$ and $\nu$ at the synchrotron characteristic frequencies ($\nu_{\rm ch}$). According to the external forward shock model, $F_{\nu} (\nu_{\rm ch})$ and $\nu_{\rm ch}$ can be described as a function of time, providing specific scaling relations \citep[e.g.,][]{Sari1998, Granot2002}. Depending on the circumburst density profile and the cooling regime, the exponent of the scaling relation is determined. If our observed break corresponds to the highest characteristic frequency among two frequencies related to the cooling or minimum electron Lorentz factors, in the slow cooling regime, the relation is $F_{\nu} (\nu_{\rm ch}) \propto {\nu_{\rm ch}}\!^{(p-1)}$ (uniform medium) or $F_{\nu} (\nu_{\rm ch}) \propto {\nu_{\rm ch}}\!^{(1-2p)}$ (wind medium), where $p$ is the electron spectral index. In the fast cooling regime, the relation is $F_{\nu} (\nu_{\rm ch}) \propto {\nu_{\rm ch}}\!^{-1/3}$, regardless of the density profile. Those relations are unlikely to produce the observed $F_{\nu} (\nu_p) \propto \nu_p^2$ relation with a reasonable value of the electron spectral index \textit{p}. 

\section{Conclusion} \label{sec:conclusion}
In this work, we selected 32 broad pulses in 32 GRBs from 175 bright GRBs by imposing reasonable selection criteria, and performed a time-resolved spectral analysis on those GRBs. We tested the HLE relation, $F_{\nu, E_{p}} \propto E_{p}\!^{2}$, and found that 18 out of 32 broad pulses exhibit consistency with the relation. The exponent $\delta$ in the HLE relation (Equation~\ref{eq:hle}) is distributed as the normal distribution with median and width of 1.99 and 0.34, respectively. The clear sample also satisfied the other two HLE relations, $t_{\rm obs}-E_{p}$ and $t_{\rm obs}-F_{\nu, E_{p}}$ (Equation~\ref{eq:add}), which supports that the identification of the HLE relation is not made accidentally. Our result agrees with the prediction of the HLE theory and successfully identifies the relation during the prompt emission for the first time. The HLE signature is not expected to be observed in all GRBs, because the HLE could be buried under LoSE depending the physical conditions in the emitting region (Uhm et al. 2022, submitted). 

Other than the HLE signature, we found several unusual features. We found $\delta$ of 0.7 in three GRBs, which can be associated with a different physical origin. Especially, GRB\,110920A shows a break in the $E_{p}$ and $F_{\nu, E_{p}}$ evolution and has two distinct $\delta$ values, 0.7 (before the break) and 2 (after the break). The later $delta$ value is consistent with the HLE prediction, whereas the former value might originate from a different physical process.  We also found that there may be a positive correlation between $\alpha$ and $F_{\nu, E_{p}}$; i.e., as $F_{\nu, E_{p}}$ decreases, $\alpha$ softens. Since this correlation is noticeably observed only in GRBs having the clear or weak HLE evidence, the correlation may be related to HLE. 

The observation of the HLE evidence in relatively long GRBs ($T_{90} \gtrsim 10 s$) implies that the emission radius for those GRBs is $r \sim 10^{16} cm$ (Uhm et al. 2022, submitted). This large emission radius of the gamma-ray emitting region disfavors some prompt emission models such as the photosphere model or the internal shock model but favors the magnetic dissipation models that invoke a large dissipation radius, such as the ICMART model \citep{Zhang2011}.

\acknowledgments
D.Tak acknowledges the Young Investigators Program of the Helmholtz Association. D.Tak and Z.L.Uhm are supported by the National Research Foundation of Korea (NRF) grant, No. 2021M3F7A1084525, funded by the Korea government (MSIT). B.B.Z acknowledges the support by the National Key Research and Development Programs of China (2018YFA0404204, 2022YFF0711404, 2022SKA0130102), the National Natural Science Foundation of China (Grant Nos. 11833003, U2038105, 12121003), the science research grants from the China Manned Space Project with NO.CMS-CSST-2021-B11, and the Program for Innovative Talents, Entrepreneur in Jiangsu.

\bibliographystyle{aasjournal}
\bibliography{references}
\appendix
\section{Table for parameters related to selection criteria}\label{sec:app1}
\begin{longtable*}{c c c c c c c c c c c c c c c c c}
\hline\hline
& & & & & & & & \multicolumn{6}{c}{Criteria\footnote{Details of criteria are in the Section~\ref{sec:selection}. The yellow and red flags are marked as $\bigtriangleup$ and X, respectiv}}\\\cline{9-14}
    Trigger name & Fluence & Peak flux & Flu. \%  & t$_{rise}$ & t$_{decay}$& t$_{bumps}$ & & 1 & 2 & 3 & 4& Final\\ 
     & [\,10$^{-5}$ erg cm$^{-2}$\,] & [\,10$^{-6}$ erg cm$^{-2}$ s$^{-1}$\,] & & [\,s\,] & [\,s\,] & [\,s\,] & & & & & &\\\hline
bn080723557 & 8.3 & 5.5 & 20 & 1.7 & 1.3 & 0.0 & & X & $\bigtriangleup$ & - & X & - \\
bn080723985 & 3.6 & 2.4 & 72 & 11.7 & 3.4 & 0.0 & & $\bigtriangleup$ & X & - & - & - \\
bn080817161 & 5.5 & 3.1 & 100 & 7.8 & 17.1 & 9.5 & & - & - & X & - & - \\
bn080825593 & 3.7 & 5.0 & 55 & 3.2 & 4.6 & 1.3 & & X & - & X & X & - \\
bn080916009 & 8.3 & 3.7 & 100 & 2.7 & 45.9 & 20.1 & & - & - & X & $\bigtriangleup$ & - \\
bn081009140 & 3.7 & 8.2 & 100 & 3.0 & 4.3 & 0.9 & & - & - & $\bigtriangleup$ & - & O \\
bn081215784 & 5.4 & 23.1 & 100 & 1.9 & 6.2 & 3.6 & & - & - & X & - & - \\
bn081221681 & 2.9 & 2.4 & 100 & 4.4 & 11.3 & 0.8 & & - & - & $\bigtriangleup$ & - & O \\
bn081224887 & 3.6 & 6.7 & 100 & 2.7 & 10.0 & 0.0 & & - & - & - & - & O \\
bn090102122 & 3.3 & 3.0 & 35 & 0.2 & 3.3 & 2.7 & & X & - & X & X & - \\
bn090217206 & 3.2 & 3.1 & 100 & 6.0 & 6.8 & 6.2 & & - & - & X & X & - \\
bn090323002 & 12.5 & 3.2 & 62 & 29.1 & 5.3 & 0.4 & & X & X & $\bigtriangleup$ & $\bigtriangleup$ & - \\
bn090328401 & 5.2 & 4.1 & 75 & 14.2 & 4.3 & 0.0 & & $\bigtriangleup$ & X & - & X & - \\
bn090424592 & 4.6 & 10.8 & 100 & 4.4 & 1.5 & 0.5 & & - & X & X & X & - \\
bn090528516 & 4.4 & 1.9 & 18 & 2.7 & 2.4 & 0.0 & & X & $\bigtriangleup$ & - & X & - \\
bn090618353 & 27.0 & 13.7 & 85 & 14.1 & 33.3 & 14.8 & & $\bigtriangleup$ & - & X & - & - \\
bn090626189 & 7.2 & 5.1 & 18 & 2.8 & 1.6 & 0.0 & & X & $\bigtriangleup$ & - & X & - \\
bn090717034 & 2.7 & 2.0 & 70 & 5.6 & 7.3 & 0.0 & & $\bigtriangleup$ & - & - & - & O \\
bn090719063 & 4.7 & 7.0 & 100 & 5.2 & 8.5 & 0.0 & & - & - & - & - & O \\
bn090820027 & 15.2 & 23.0 & 100 & 5.2 & 10.4 & 2.3 & & - & - & $\bigtriangleup$ & - & O \\
bn090829672 & 9.1 & 6.5 & 89 & 13.0 & 14.4 & 3.8 & & $\bigtriangleup$ & - & X & - & - \\
bn090902462 & 27.9 & 19.9 & 99 & 15.2 & 8.4 & 5.3 & & - & $\bigtriangleup$ & X & - & - \\
bn090926181 & 15.3 & 18.6 & 100 & 4.5 & 13.4 & 7.9 & & - & - & X & - & - \\
bn091003191 & 3.6 & 8.5 & 60 & 2.5 & 3.4 & 0.9 & & X & - & X & X & - \\
bn091030828 & 3.3 & 3.3 & 51 & 1.5 & 8.2 & 0.0 & & X & - & - & X & - \\
bn100116897 & 3.7 & 4.6 & 94 & 7.2 & 5.4 & 0.0 & & - & $\bigtriangleup$ & - & X & - \\
bn100322045 & 6.3 & 2.2 & 68 & 14.0 & 6.7 & 0.0 & & X & X & - & X & - \\
bn100324172 & 4.5 & 5.3 & 100 & 5.0 & 6.8 & 0.0 & & - & - & - & - & O \\
bn100414097 & 9.2 & 5.5 & 100 & 24.2 & 1.8 & 0.0 & & - & X & - & - & - \\
bn100511035 & 3.1 & 3.5 & 63 & 2.1 & 6.3 & 0.8 & & X & - & X & X & - \\
bn100528075 & 3.0 & 2.5 & 94 & 8.0 & 7.5 & 1.0 & & - & $\bigtriangleup$ & $\bigtriangleup$ & X & - \\
bn100701490 & 2.9 & 8.8 & 14 & 0.3 & 0.3 & 0.1 & & X & $\bigtriangleup$ & X & X & - \\
bn100719989 & 5.1 & 15.2 & 88 & 1.4 & 4.4 & 1.6 & & $\bigtriangleup$ & - & X & - & - \\
bn100724029 & 24.3 & 6.5 & 97 & 61.0 & 23.8 & 5.9 & & - & X & X & - & - \\
bn100728095 & 12.0 & 3.2 & 72 & 37.2 & 11.5 & 4.4 & & $\bigtriangleup$ & X & X & X & - \\
bn100826957 & 18.2 & 8.2 & 70 & 21.5 & 25.7 & 4.6 & & X & - & X & - & - \\
bn100906576 & 2.6 & 2.1 & 100 & 11.0 & 2.0 & 0.0 & & - & X & - & - & - \\
bn100918863 & 14.9 & 3.7 & 67 & 46.0 & 12.9 & 0.0 & & X & X & - & $\bigtriangleup$ & - \\
bn101014175 & 17.9 & 12.7 & 33 & 1.8 & 9.3 & 7.2 & & X & - & X & X & - \\
\hline
\caption{Selection criteria for 175 GRBs}
\\\\
\hline
bn101023951 & 5.8 & 5.6 & 100 & 4.6 & 14.4 & 2.0 & & - & - & $\bigtriangleup$ & - & O \\
bn101123952 & 12.0 & 9.6 & 81 & 11.4 & 9.4 & 4.0 & & $\bigtriangleup$ & $\bigtriangleup$ & X & - & - \\
bn110102788 & 4.1 & 2.8 & 65 & 11.7 & 3.6 & 0.0 & & X & X & - & - & - \\
bn110123804 & 2.6 & 1.9 & 86 & 10.5 & 5.2 & 0.0 & & $\bigtriangleup$ & X & - & X & - \\
bn110301214 & 3.8 & 10.7 & 100 & 4.0 & 2.9 & 0.0 & & - & $\bigtriangleup$ & - & - & O \\
bn110407998 & 2.9 & 4.6 & 100 & 3.5 & 6.6 & 1.8 & & - & - & X & - & - \\
bn110622158 & 5.5 & 1.9 & 100 & 15.5 & 26.1 & 7.0 & & - & - & X & - & - \\
bn110625881 & 7.0 & 12.2 & 61 & 3.2 & 6.8 & 2.4 & & X & - & X & - & - \\
bn110709642 & 4.3 & 3.6 & 11 & 2.2 & 0.6 & 0.0 & & X & X & - & X & - \\
bn110717319 & 4.8 & 2.3 & 85 & 12.9 & 3.8 & 2.0 & & $\bigtriangleup$ & X & X & - & - \\
bn110721200 & 3.9 & 6.0 & 100 & 2.4 & 9.0 & 0.0 & & - & - & - & - & O \\
bn110729142 & 5.6 & 1.8 & 100 & 10.3 & 4.3 & 1.3 & & - & X & X & - & - \\
bn110825102 & 4.7 & 12.0 & 100 & 4.4 & 4.5 & 2.6 & & - & - & X & - & - \\
bn110919634 & 2.7 & 1.8 & 100 & 14.6 & 10.1 & 1.6 & & - & $\bigtriangleup$ & $\bigtriangleup$ & - & - \\
bn110920546 & 15.9 & 4.3 & 100 & 10.9 & 78.1 & 0.0 & & - & - & - & - & O \\
bn110921912 & 3.7 & 7.9 & 83 & 2.6 & 5.4 & 1.5 & & $\bigtriangleup$ & - & X & $\bigtriangleup$ & - \\
bn111220486 & 5.7 & 6.1 & 65 & 7.6 & 7.2 & 1.4 & & X & $\bigtriangleup$ & $\bigtriangleup$ & X & - \\
bn120119170 & 3.9 & 3.0 & 100 & 11.4 & 17.7 & 6.1 & & - & - & X & - & - \\
bn120129580 & 5.7 & 32.2 & 100 & 1.6 & 3.2 & 0.8 & & - & - & X & - & - \\
bn120204054 & 9.5 & 4.4 & 95 & 13.1 & 32.1 & 3.7 & & - & - & $\bigtriangleup$ & - & O \\
bn120226871 & 6.1 & 2.0 & 100 & 16.8 & 22.3 & 16.0 & & - & - & X & - & - \\
bn120316008 & 2.5 & 1.9 & 23 & 1.2 & 0.7 & 0.0 & & X & $\bigtriangleup$ & - & X & - \\
bn120328268 & 8.3 & 7.2 & 100 & 5.9 & 28.1 & 12.7 & & - & - & X & - & - \\
bn120526303 & 13.4 & 3.2 & 78 & 1.8 & 32.9 & 31.9 & & $\bigtriangleup$ & - & X & X & - \\
bn120624933 & 19.6 & 5.2 & 100 & 9.6 & 7.0 & 0.0 & & - & $\bigtriangleup$ & - & - & O \\
bn120707800 & 10.3 & 7.4 & 100 & 27.2 & 11.0 & 0.0 & & - & X & - & - & - \\
bn120711115 & 19.7 & 8.0 & 100 & 33.1 & 12.3 & 4.9 & & - & X & X & - & - \\
bn120728434 & 12.8 & 4.2 & 36 & 2.6 & 19.8 & 6.9 & & X & - & X & - & - \\
bn121122885 & 5.3 & 7.2 & 100 & 1.7 & 7.4 & 1.3 & & - & - & $\bigtriangleup$ & - & O \\
bn121225417 & 7.3 & 5.1 & 16 & 1.9 & 2.2 & 1.0 & & X & - & X & X & - \\
bn130121835 & 4.6 & 3.9 & 100 & 6.3 & 12.4 & 0.7 & & - & - & $\bigtriangleup$ & X & - \\
bn130219775 & 3.3 & 3.0 & 91 & 7.8 & 12.4 & 0.0 & & - & - & - & - & O \\
bn130304410 & 4.7 & 3.3 & 86 & 3.3 & 8.6 & 7.7 & & $\bigtriangleup$ & - & X & X & - \\
bn130305486 & 6.0 & 10.1 & 100 & 5.9 & 8.9 & 0.0 & & - & - & - & $\bigtriangleup$ & O \\
bn130306991 & 16.8 & 4.9 & 100 & 11.4 & 12.7 & 7.0 & & - & - & X & $\bigtriangleup$ & - \\
bn130327350 & 5.3 & 2.7 & 28 & 1.2 & 6.5 & 3.2 & & X & - & X & X & - \\
bn130425327 & 5.2 & 3.4 & 27 & 0.7 & 1.6 & 0.4 & & X & - & X & X & - \\
bn130427324 & 141.2 & 176.0 & 93 & 5.6 & 60.4 & 13.7 & & - & - & X & - & - \\
bn130502327 & 11.4 & 11.0 & 14 & 2.2 & 1.0 & 0.0 & & X & X & - & X & - \\
bn130504978 & 13.3 & 10.9 & 64 & 22.7 & 4.5 & 0.0 & & X & X & - & - & - \\
bn130518580 & 10.0 & 11.8 & 100 & 11.9 & 11.0 & 0.9 & & - & $\bigtriangleup$ & $\bigtriangleup$ & - & - \\
bn130606497 & 20.9 & 17.2 & 57 & 9.2 & 13.1 & 3.9 & & X & - & X & X & - \\
bn130609902 & 6.5 & 3.5 & 100 & 7.7 & 19.7 & 6.4 & & - & - & X & - & - \\
bn130704560 & 2.5 & 5.5 & 100 & 4.0 & 2.0 & 0.3 & & - & $\bigtriangleup$ & $\bigtriangleup$ & - & - \\
bn130715906 & 4.4 & 2.5 & 100 & 23.1 & 15.3 & 2.9 & & - & $\bigtriangleup$ & X & - & - \\
bn130720582 & 10.3 & 2.2 & 69 & 35.6 & 13.9 & 2.1 & & X & X & $\bigtriangleup$ & $\bigtriangleup$ & - \\
bn130821674 & 7.1 & 4.9 & 69 & 9.5 & 7.5 & 1.5 & & X & $\bigtriangleup$ & $\bigtriangleup$ & X & - \\
\hline
\caption{Selection criteria for 175 GRBs (\textit{continue})}
\\\\
\hline
bn131014215 & 18.7 & 87.1 & 100 & 1.7 & 4.3 & 1.8 & & - & - & X & - & - \\
bn131028076 & 16.1 & 28.4 & 100 & 5.8 & 10.9 & 0.0 & & - & - & - & - & O \\
bn131108862 & 3.6 & 4.5 & 20 & 0.3 & 1.6 & 1.1 & & X & - & X & X & - \\
bn131118958 & 8.0 & 3.2 & 9 & 1.3 & 0.3 & 0.0 & & X & X & - & X & - \\
bn131122490 & 3.5 & 3.4 & 35 & 0.7 & 3.8 & 1.8 & & X & - & X & X & - \\
bn131127592 & 4.0 & 4.1 & 78 & 5.5 & 8.5 & 1.6 & & $\bigtriangleup$ & - & X & X & - \\
bn131214705 & 6.5 & 3.0 & 76 & 7.4 & 14.5 & 0.0 & & $\bigtriangleup$ & - & - & - & O \\
bn131229277 & 2.7 & 7.7 & 75 & 3.0 & 2.1 & 0.8 & & $\bigtriangleup$ & $\bigtriangleup$ & X & X & - \\
bn131231198 & 15.3 & 14.8 & 100 & 21.4 & 25.6 & 6.8 & & - & - & X & - & - \\
bn140206275 & 12.8 & 9.8 & 97 & 8.6 & 29.7 & 5.4 & & - & - & $\bigtriangleup$ & - & O \\
bn140306146 & 7.4 & 5.9 & 56 & 1.9 & 8.2 & 7.6 & & X & - & X & X & - \\
bn140323433 & 3.0 & 2.1 & 20 & 1.7 & 1.2 & 0.0 & & X & $\bigtriangleup$ & - & X & - \\
bn140329295 & 7.2 & 21.5 & 95 & 4.7 & 4.5 & 0.0 & & - & $\bigtriangleup$ & - & - & O \\
bn140416060 & 7.9 & 8.2 & 53 & 7.2 & 3.5 & 1.2 & & X & X & X & X & - \\

bn140508128 & 6.4 & 12.3 & 60 & 2.7 & 5.1 & 0.0 & & X & - & - & X & - \\
bn140512814 & 4.5 & 2.5 & 52 & 6.4 & 10.3 & 4.7 & & X & - & X & X & - \\
bn140523129 & 5.3 & 7.2 & 17 & 1.0 & 1.6 & 0.0 & & X & - & - & X & - \\
bn140721336 & 5.1 & 2.4 & 56 & 6.0 & 2.1 & 0.0 & & X & X & - & X & - \\
bn140810782 & 11.8 & 5.2 & 29 & 6.8 & 3.1 & 0.0 & & X & X & - & X & - \\
bn140821997 & 6.4 & 4.8 & 100 & 20.0 & 13.0 & 1.9 & & - & $\bigtriangleup$ & X & - & - \\
bn141022087 & 9.5 & 18.1 & 100 & 5.7 & 3.0 & 1.7 & & - & $\bigtriangleup$ & X & X & - \\
bn141028455 & 4.0 & 4.3 & 100 & 7.2 & 12.2 & 0.8 & & - & - & $\bigtriangleup$ & - & O \\
bn141029134 & 4.2 & 4.5 & 58 & 3.7 & 3.9 & 1.2 & & X & - & X & X & - \\
bn141207800 & 3.7 & 4.5 & 76 & 6.8 & 3.1 & 1.8 & & $\bigtriangleup$ & X & X & X & - \\
bn141215560 & 2.9 & 4.7 & 30 & 0.6 & 1.3 & 0.3 & & X & - & $\bigtriangleup$ & X & - \\
bn141222691 & 2.6 & 3.1 & 60 & 3.9 & 4.7 & 0.0 & & X & - & - & X & - \\
bn150118409 & 14.9 & 13.1 & 21 & 2.9 & 3.3 & 0.4 & & X & - & X & X & - \\
bn150201574 & 6.7 & 9.5 & 100 & 1.9 & 16.0 & 6.1 & & - & - & X & - & - \\
bn150202999 & 3.5 & 4.4 & 100 & 4.8 & 8.3 & 4.0 & & - & - & X & - & - \\
bn150210935 & 3.0 & 10.8 & 92 & 1.3 & 2.2 & 1.0 & & - & - & X & X & - \\
bn150213001 & 3.0 & 12.7 & 100 & 2.3 & 3.1 & 0.0 & & - & - & - & - & O \\
bn150220598 & 2.8 & 2.4 & 47 & 2.2 & 1.8 & 0.0 & & X & $\bigtriangleup$ & - & X & - \\
bn150306993 & 3.0 & 4.0 & 100 & 3.9 & 9.3 & 0.0 & & - & - & - & - & O \\
bn150309958 & 4.2 & 2.5 & 100 & 16.4 & 11.3 & 2.0 & & - & $\bigtriangleup$ & $\bigtriangleup$ & - & - \\
bn150314205 & 8.9 & 16.0 & 100 & 1.4 & 11.0 & 1.5 & & - & - & X & - & - \\
bn150330828 & 14.7 & 10.4 & 89 & 6.9 & 24.6 & 12.0 & & $\bigtriangleup$ & - & X & - & - \\
bn150403913 & 6.4 & 9.1 & 100 & 10.5 & 13.5 & 0.0 & & - & - & - & - & O \\
bn150510139 & 9.9 & 12.9 & 90 & 0.2 & 28.3 & 23.4 & & $\bigtriangleup$ & - & X & X & - \\
bn150523396 & 3.7 & 3.6 & 55 & 6.2 & 2.9 & 0.0 & & X & X & - & - & - \\
bn150627183 & 18.5 & 10.5 & 58 & 27.4 & 13.9 & 6.7 & & X & $\bigtriangleup$ & X & - & - \\
bn150724782 & 3.6 & 3.0 & 88 & 11.3 & 3.2 & 0.8 & & $\bigtriangleup$ & X & X & X & - \\
bn150821406 & 7.1 & 2.1 & 100 & 15.2 & 42.5 & 12.4 & & - & - & X & - & - \\
bn150902733 & 8.7 & 13.5 & 100 & 9.3 & 11.7 & 0.8 & & - & - & $\bigtriangleup$ & - & O \\
bn151030999 & 5.5 & 3.1 & 89 & 14.0 & 15.0 & 12.0 & & $\bigtriangleup$ & - & X & - & - \\
bn151107851 & 2.8 & 3.7 & 100 & 8.6 & 11.6 & 0.0 & & - & - & - & - & O \\
bn151227218 & 4.5 & 7.6 & 78 & 7.1 & 7.9 & 1.3 & & $\bigtriangleup$ & - & X & X & - \\
bn151231443 & 9.4 & 9.8 & 53 & 6.2 & 9.5 & 0.6 & & X & - & $\bigtriangleup$ & X & - \\
\hline
\caption{Selection criteria for 175 GRBs (\textit{continue})}
\\\\
\hline
bn160106948 & 5.1 & 6.1 & 71 & 3.3 & 11.9 & 5.0 & & $\bigtriangleup$ & - & X & - & - \\
bn160107931 & 2.9 & 2.3 & 38 & 0.3 & 3.8 & 0.8 & & X & - & $\bigtriangleup$ & X & - \\
bn160113398 & 3.5 & 3.1 & 100 & 7.9 & 9.5 & 0.0 & & - & - & - & - & O \\
bn160118060 & 3.5 & 3.2 & 23 & 0.6 & 1.2 & 0.0 & & X & - & - & X & - \\
bn160215773 & 6.4 & 4.2 & 100 & 17.8 & 14.2 & 4.3 & & - & $\bigtriangleup$ & X & - & - \\
bn160421137 & 4.3 & 2.8 & 100 & 12.3 & 4.3 & 1.1 & & - & X & X & - & - \\
bn160422499 & 9.0 & 17.8 & 100 & 8.4 & 5.5 & 0.7 & & - & $\bigtriangleup$ & $\bigtriangleup$ & - & - \\
bn160509374 & 20.4 & 14.9 & 92 & 14.6 & 20.1 & 2.6 & & - & - & $\bigtriangleup$ & - & O \\
bn160530667 & 9.8 & 18.4 & 100 & 6.4 & 9.8 & 0.4 & & - & - & $\bigtriangleup$ & - & O \\
bn160625945 & 66.8 & 66.6 & 99 & 4.4 & 33.5 & 18.1 & & - & - & X & - & - \\
bn160720767 & 15.4 & 11.2 & 100 & 11.8 & 17.1 & 3.3 & & - & - & $\bigtriangleup$ & X & - \\
bn160802259 & 6.3 & 17.6 & 81 & 1.2 & 5.1 & 1.7 & & $\bigtriangleup$ & - & X & $\bigtriangleup$ & - \\
bn160816730 & 3.4 & 6.1 & 68 & 1.1 & 3.6 & 2.8 & & X & - & X & X & - \\
bn160821857 & 54.8 & 33.4 & 100 & 25.7 & 45.2 & 4.9 & & - & - & X & - & - \\
bn160905471 & 11.2 & 6.1 & 92 & 18.7 & 5.0 & 0.3 & & - & X & $\bigtriangleup$ & $\bigtriangleup$ & - \\
bn160910722 & 8.6 & 20.1 & 100 & 8.4 & 14.2 & 0.0 & & - & - & - & - & O \\
bn161020759 & 3.1 & 4.1 & 82 & 3.5 & 7.2 & 1.2 & & $\bigtriangleup$ & - & $\bigtriangleup$ & X & - \\
bn161206064 & 3.9 & 3.5 & 92 & 8.4 & 9.6 & 1.8 & & - & - & X & X & - \\
bn161218356 & 8.8 & 9.4 & 81 & 2.2 & 19.5 & 17.1 & & $\bigtriangleup$ & - & X & X & - \\
bn161229878 & 4.2 & 3.1 & 29 & 2.9 & 3.4 & 2.2 & & X & - & X & X & - \\
bn170115743 & 7.5 & 8.9 & 29 & 1.4 & 1.3 & 0.5 & & X & $\bigtriangleup$ & X & X & - \\
bn170121614 & 3.5 & 2.4 & 59 & 9.7 & 5.1 & 0.0 & & X & $\bigtriangleup$ & - & X & - \\
bn170207906 & 6.0 & 7.5 & 49 & 2.0 & 4.2 & 1.7 & & X & - & X & X & - \\
bn170210116 & 11.5 & 7.3 & 100 & 7.3 & 29.2 & 16.5 & & - & - & X & X & - \\
bn170214649 & 19.8 & 4.4 & 90 & 46.7 & 31.1 & 24.3 & & - & $\bigtriangleup$ & X & X & - \\
bn170228794 & 2.6 & 3.5 & 100 & 1.4 & 6.4 & 0.0 & & - & - & - & X & - \\
bn170405777 & 8.1 & 3.5 & 78 & 30.3 & 26.6 & 12.5 & & $\bigtriangleup$ & $\bigtriangleup$ & X & $\bigtriangleup$ & - \\
bn170409112 & 31.9 & 29.1 & 100 & 15.7 & 36.5 & 18.5 & & - & - & X & - & - \\
bn170510217 & 5.5 & 3.5 & 100 & 16.4 & 6.2 & 1.0 & & - & X & X & - & - \\
bn170511249 & 3.3 & 2.5 & 85 & 2.8 & 12.5 & 3.9 & & $\bigtriangleup$ & - & X & X & - \\
bn170527480 & 9.2 & 8.1 & 63 & 1.2 & 14.8 & 13.1 & & X & - & X & X & - \\
bn170607946 & 5.9 & 3.9 & 100 & 19.5 & 1.6 & 0.0 & & - & X & - & - & - \\
bn170614486 & 2.7 & 2.4 & 100 & 10.7 & 1.2 & 0.0 & & - & X & - & - & - \\
bn170808936 & 11.9 & 23.7 & 100 & 16.9 & 8.9 & 1.6 & & - & $\bigtriangleup$ & X & - & - \\
bn170826819 & 3.6 & 6.2 & 100 & 2.5 & 10.1 & 7.7 & & - & - & X & - & - \\
bn170906030 & 11.2 & 4.2 & 100 & 34.1 & 43.0 & 22.3 & & - & - & X & - & - \\
bn170921168 & 7.0 & 3.3 & 100 & 2.0 & 19.6 & 1.9 & & - & - & $\bigtriangleup$ & - & O \\
bn171010792 & 67.2 & 20.1 & 97 & 27.6 & 83.2 & 60.0 & & - & - & X & - & - \\
bn171102107 & 3.2 & 4.3 & 74 & 0.4 & 8.5 & 5.7 & & $\bigtriangleup$ & - & X & X & - \\
bn171119992 & 4.9 & 8.9 & 75 & 2.3 & 3.0 & 1.0 & & $\bigtriangleup$ & - & X & X & - \\
bn171210493 & 8.2 & 3.6 & 100 & 6.7 & 47.4 & 0.0 & & - & - & - & - & O \\
bn171227000 & 30.5 & 35.6 & 100 & 20.1 & 26.5 & 12.0 & & - & - & X & - & - \\
\hline
\caption{Selection criteria for 175 GRBs (\textit{continue})}
\label{tab:samples}
\end{longtable*}

\pagebreak

\section{Spectral analysis result}\label{sec:app2}
In this section, we present the spectral analysis results of 32 pulses in 32 GRBs (see Section~\ref{sec:analysis} and Section~\ref{sec:results} for details). For each GRB, we make a figure with six panels. The color gradient used in all panels indicates the elapsed time within the time interval shaded in pink in the upper--left panel, and later time data is painted with a more saturated color. The first panel (upper--left) shows the light curves in three energy bands, with the temporal evolution of $E_{peak}$ points painted with the gradient. The second (upper--middle) and third (upper--right) panels show $\alpha$ versus $E_{p}$ and $\alpha$ versus $F_{\nu, E_{p}}$, respectively. The three lower panels are intended to test the HLE expectation with three different perspectives, $F_{\nu, E_{p}} \propto E_{p}^2$, $E_{p} \propto t^{-1}$, and $F_{\nu, E_{p}} \propto t^{-2}$ from left to right. 

\let\clearpage\relax
\begin{figure}[b]
\centering
\includegraphics[scale=0.45]{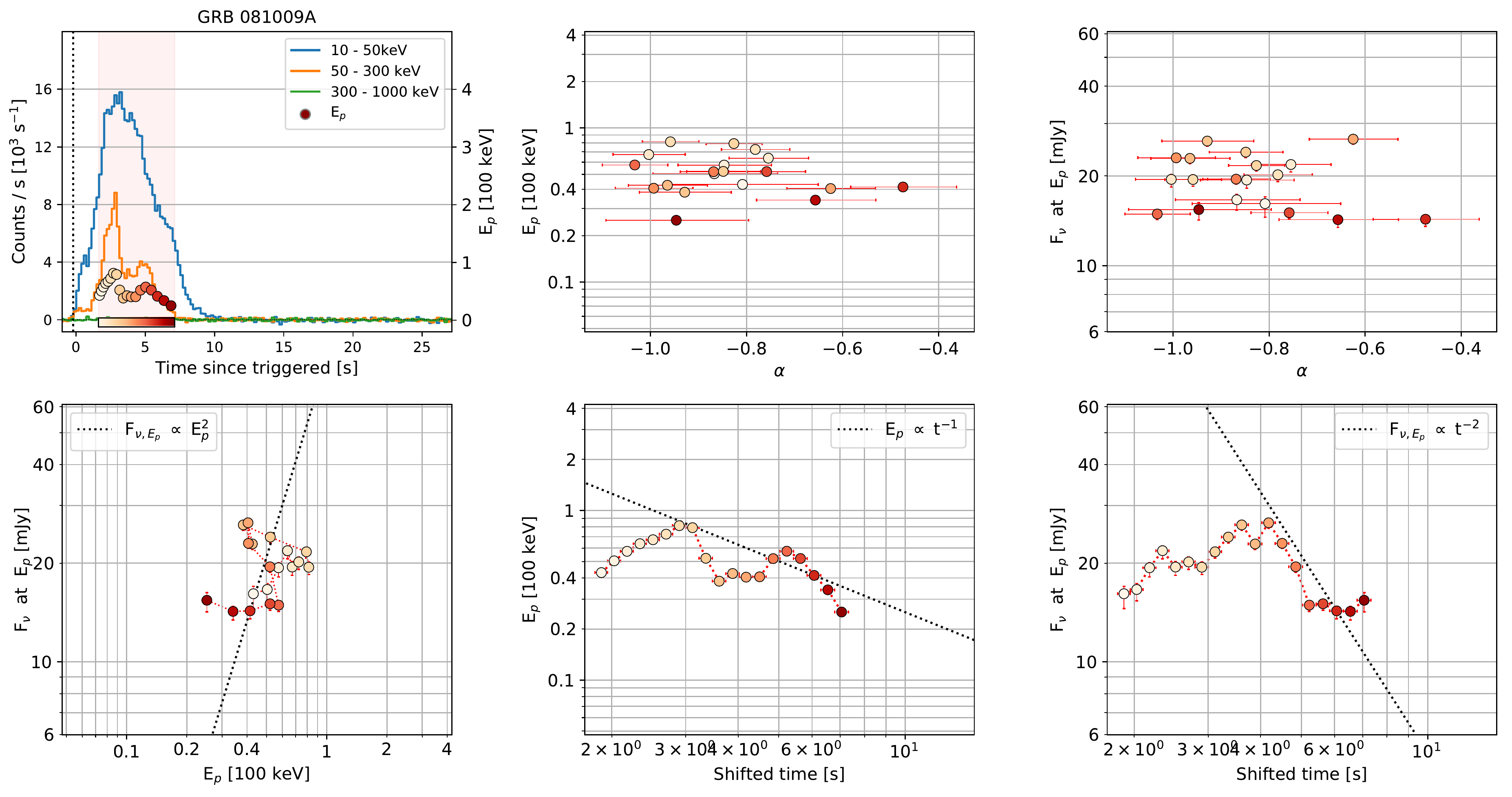} \\ 
\caption{Spectral analysis on GRB\,081009A}
\label{fig:081009A}
\vspace{2cm}
\includegraphics[scale=0.45]{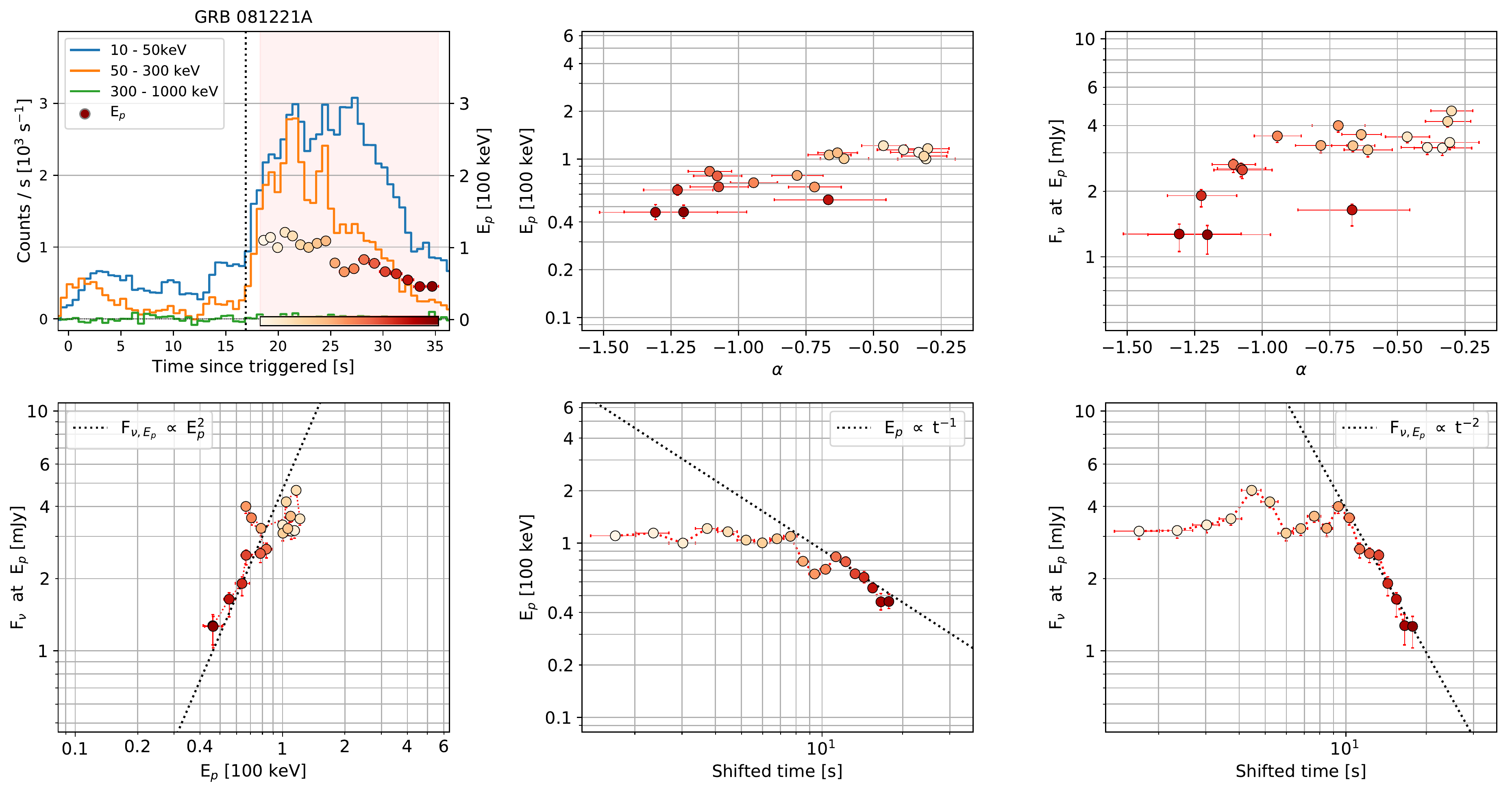} \\ 
\caption{Spectral analysis on GRB\,081221A}
\label{fig:081221A}
\end{figure}
\pagebreak
\begin{figure}
\centering
\includegraphics[scale=0.45]{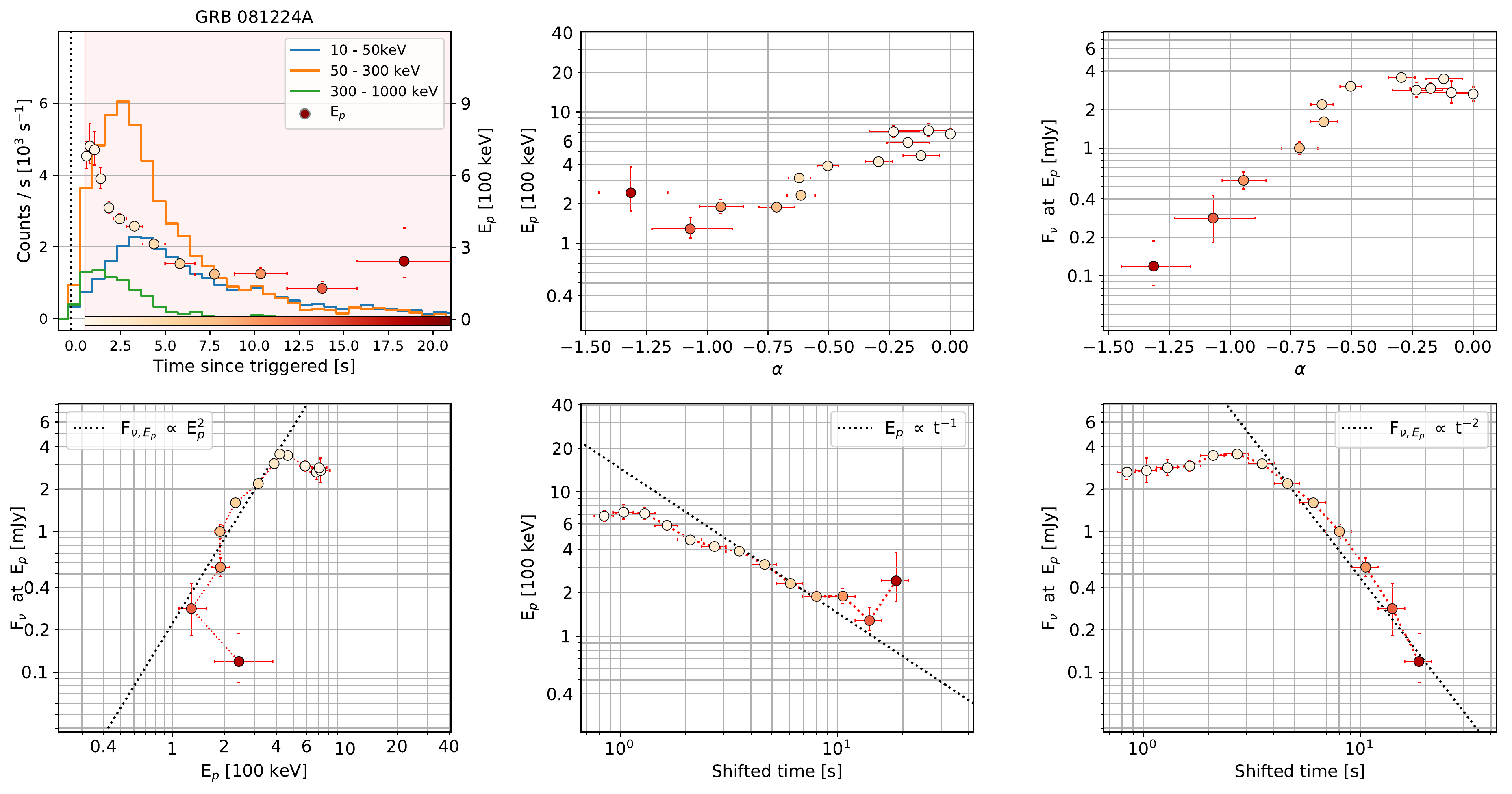} \\ 
\caption{Spectral analysis on GRB\,081224A}
\label{fig:081224A}
\vspace{2cm}
\includegraphics[scale=0.45]{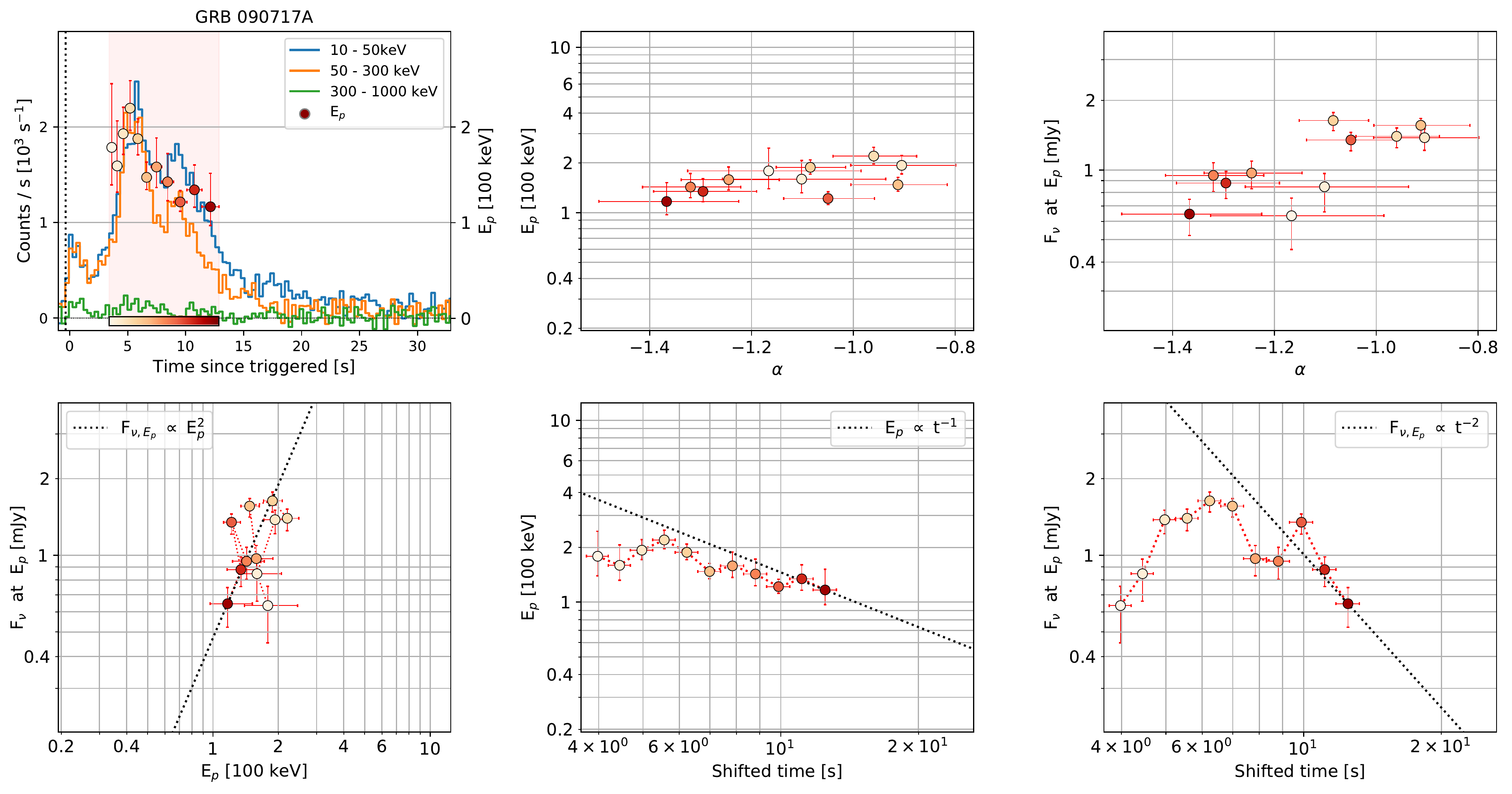} \\ 
\caption{Spectral analysis on GRB\,090717A}
\label{fig:090717A}
\end{figure}
\pagebreak
\begin{figure}
\centering
\includegraphics[scale=0.45]{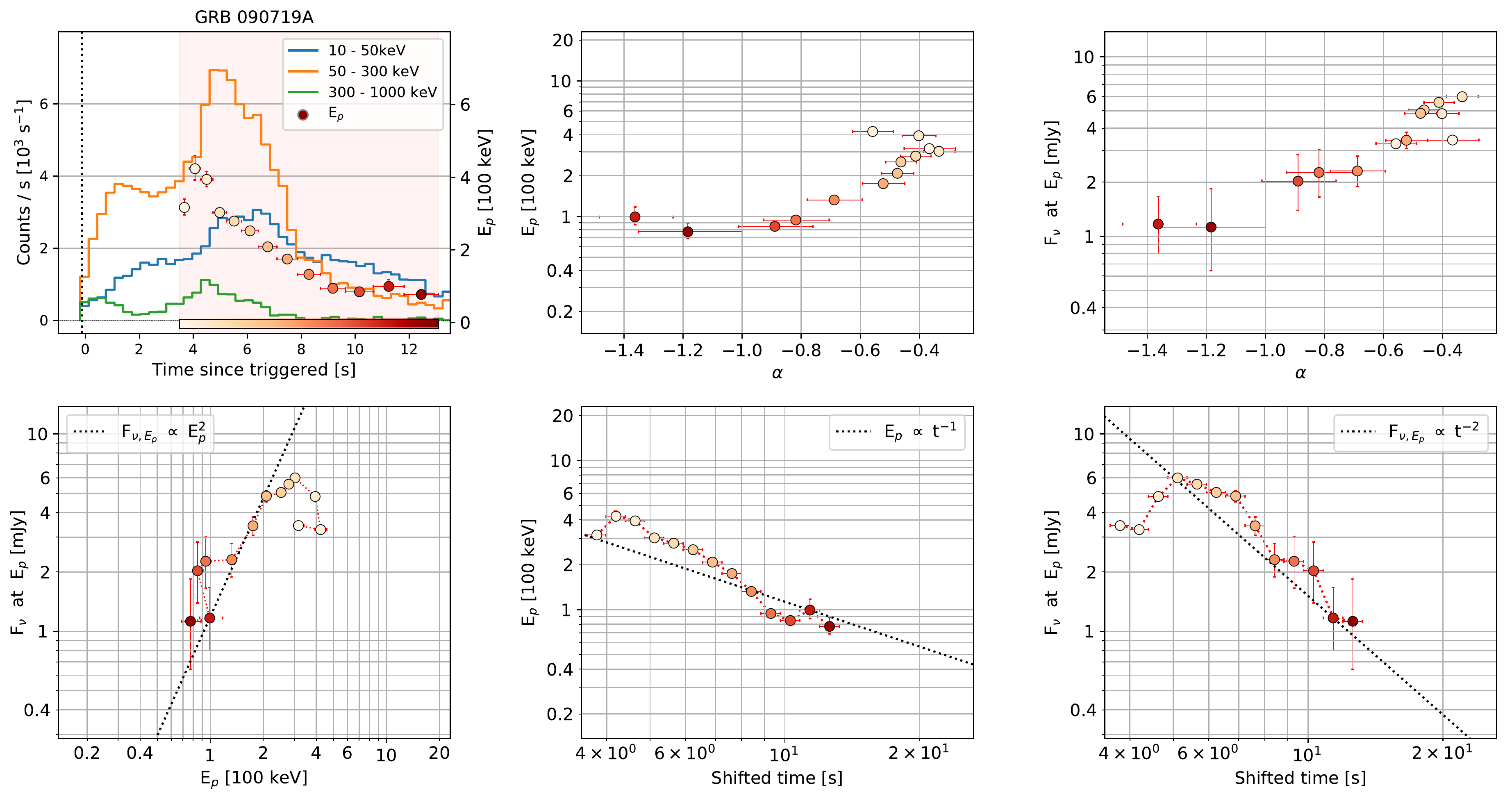} \\ 
\caption{Spectral analysis on GRB\,090719A}
\label{fig:090719A}
\vspace{2cm}
\includegraphics[scale=0.45]{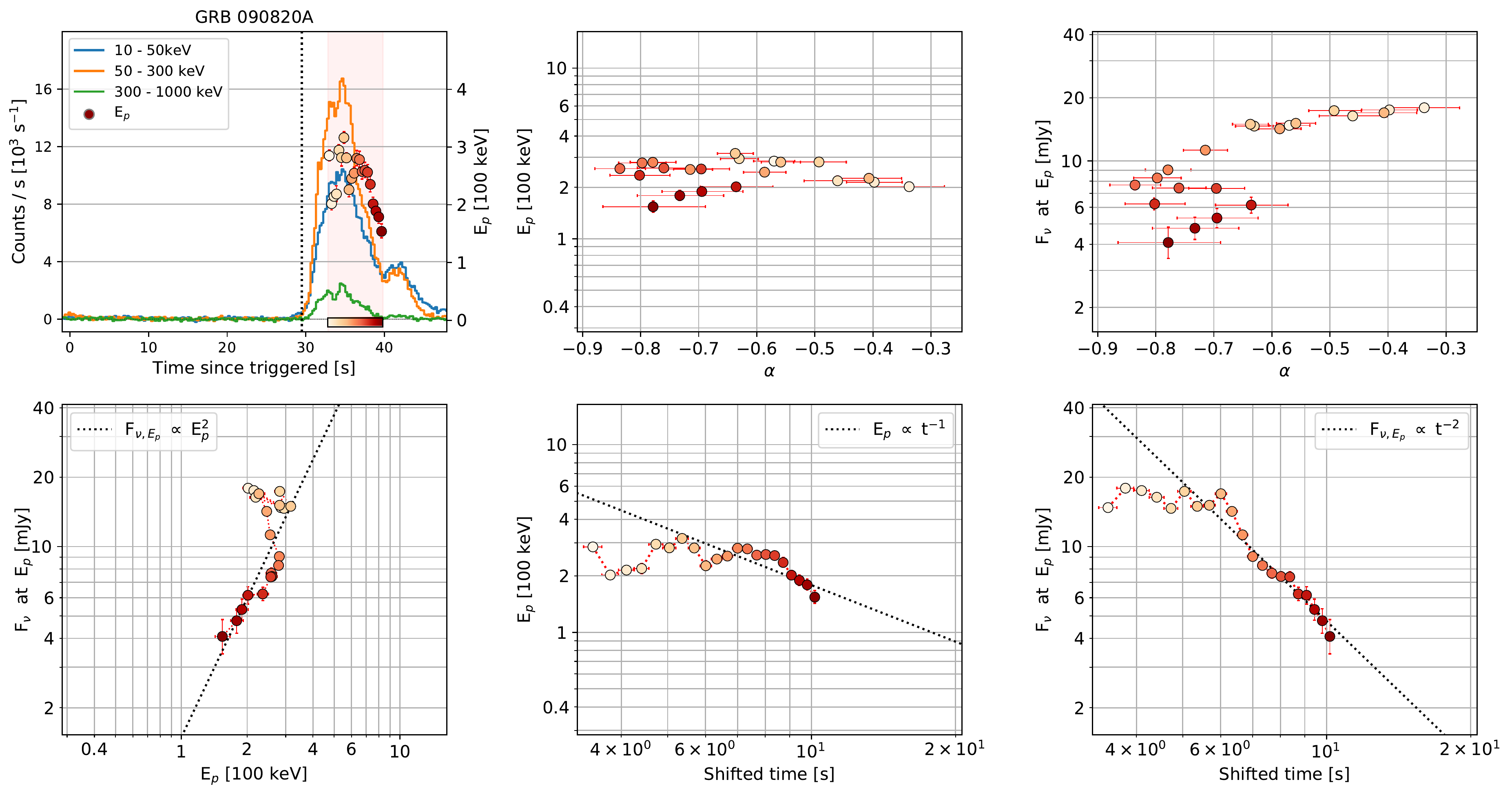} \\ 
\caption{Spectral analysis on GRB\,090820A}
\label{fig:090820A}
\end{figure}
\pagebreak
\begin{figure}
\centering
\includegraphics[scale=0.45]{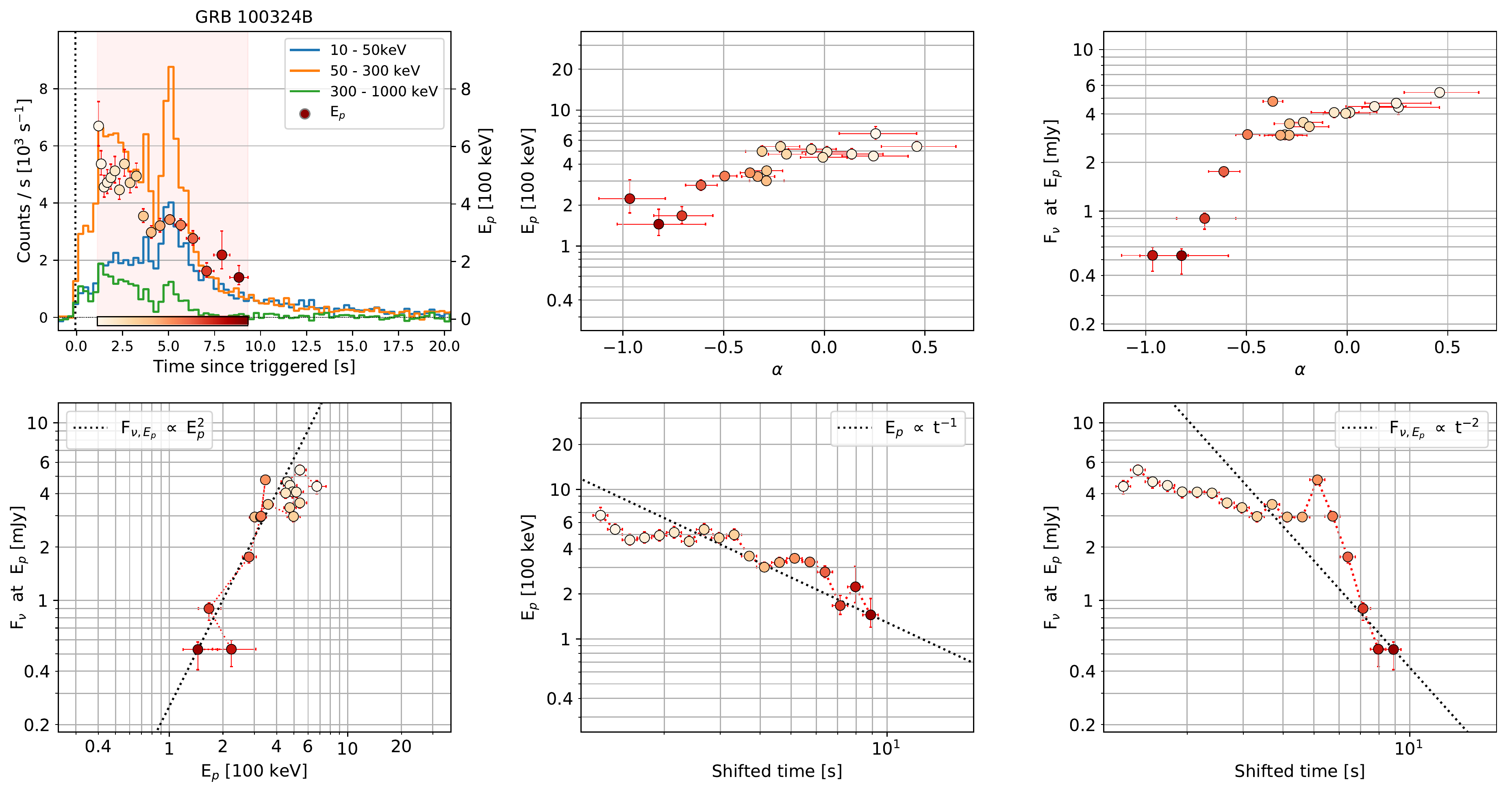} \\ 
\caption{Spectral analysis on GRB\,100324B}
\label{fig:100324B}
\vspace{2cm}
\includegraphics[scale=0.45]{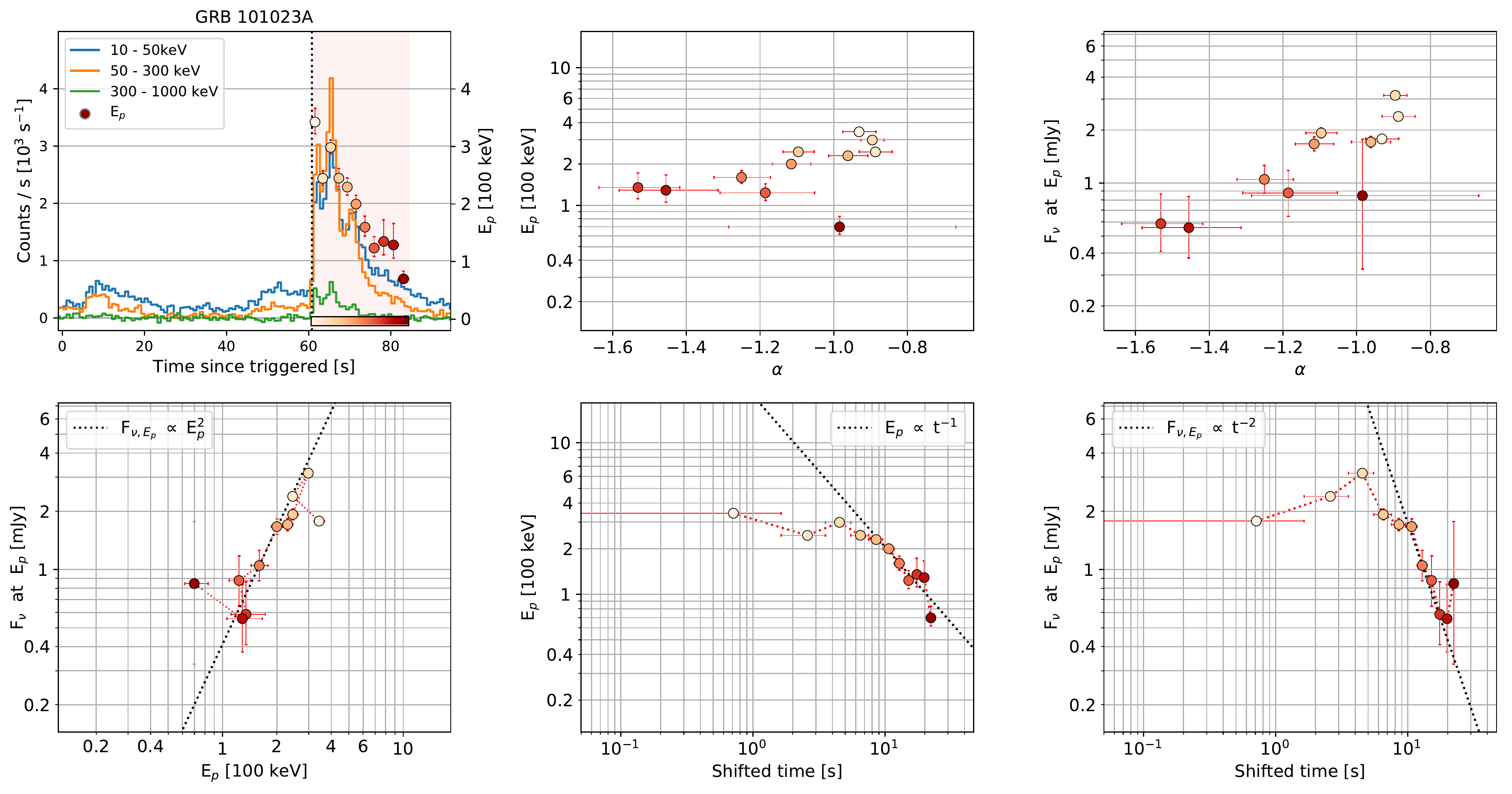} \\ 
\caption{Spectral analysis on GRB\,101023A}
\label{fig:101023A}
\end{figure}
\pagebreak
\begin{figure}
\centering
\includegraphics[scale=0.45]{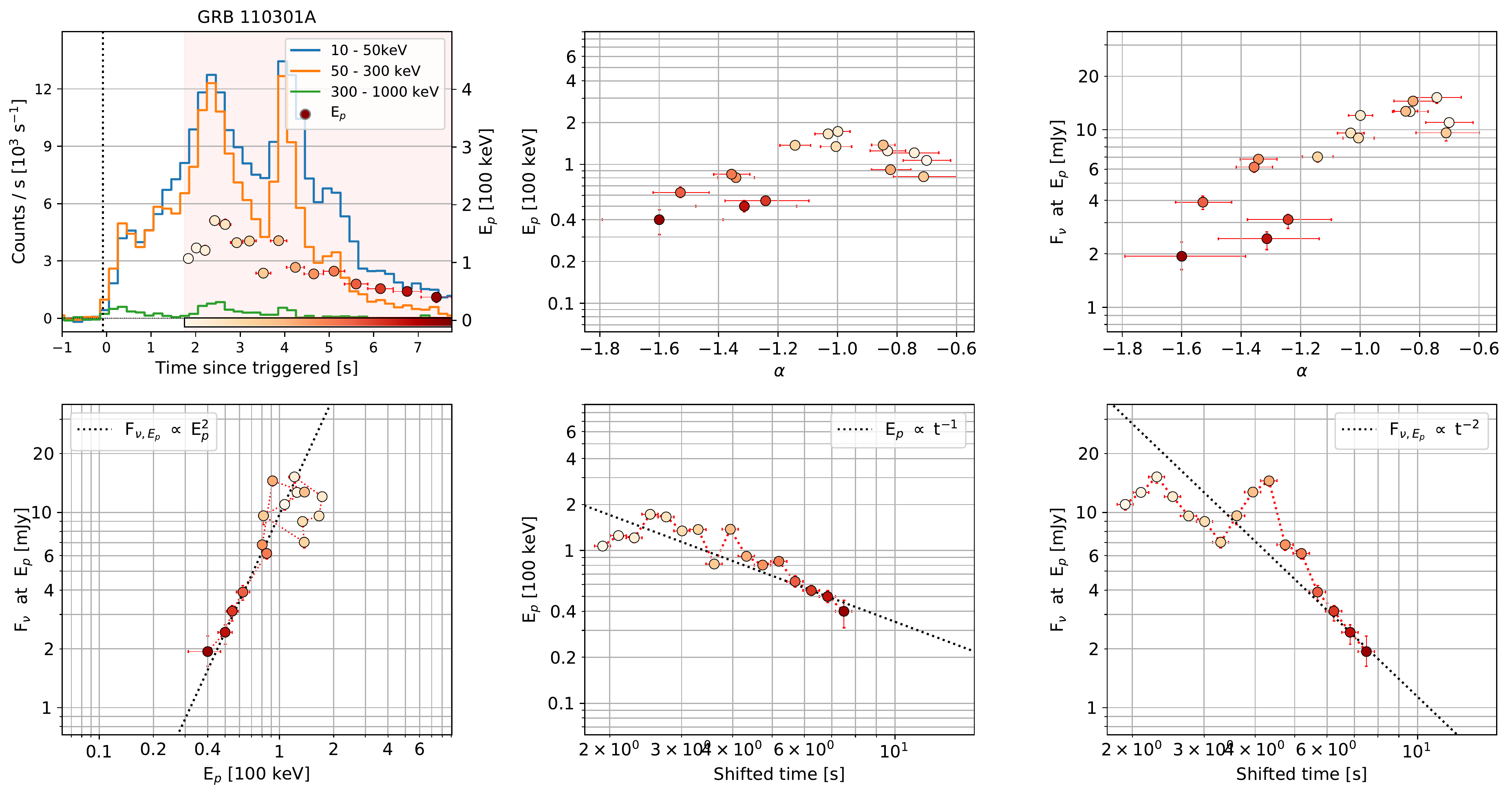} \\ 
\caption{Spectral analysis on GRB\,110301A}
\label{fig:110301A}
\vspace{2cm}
\includegraphics[scale=0.45]{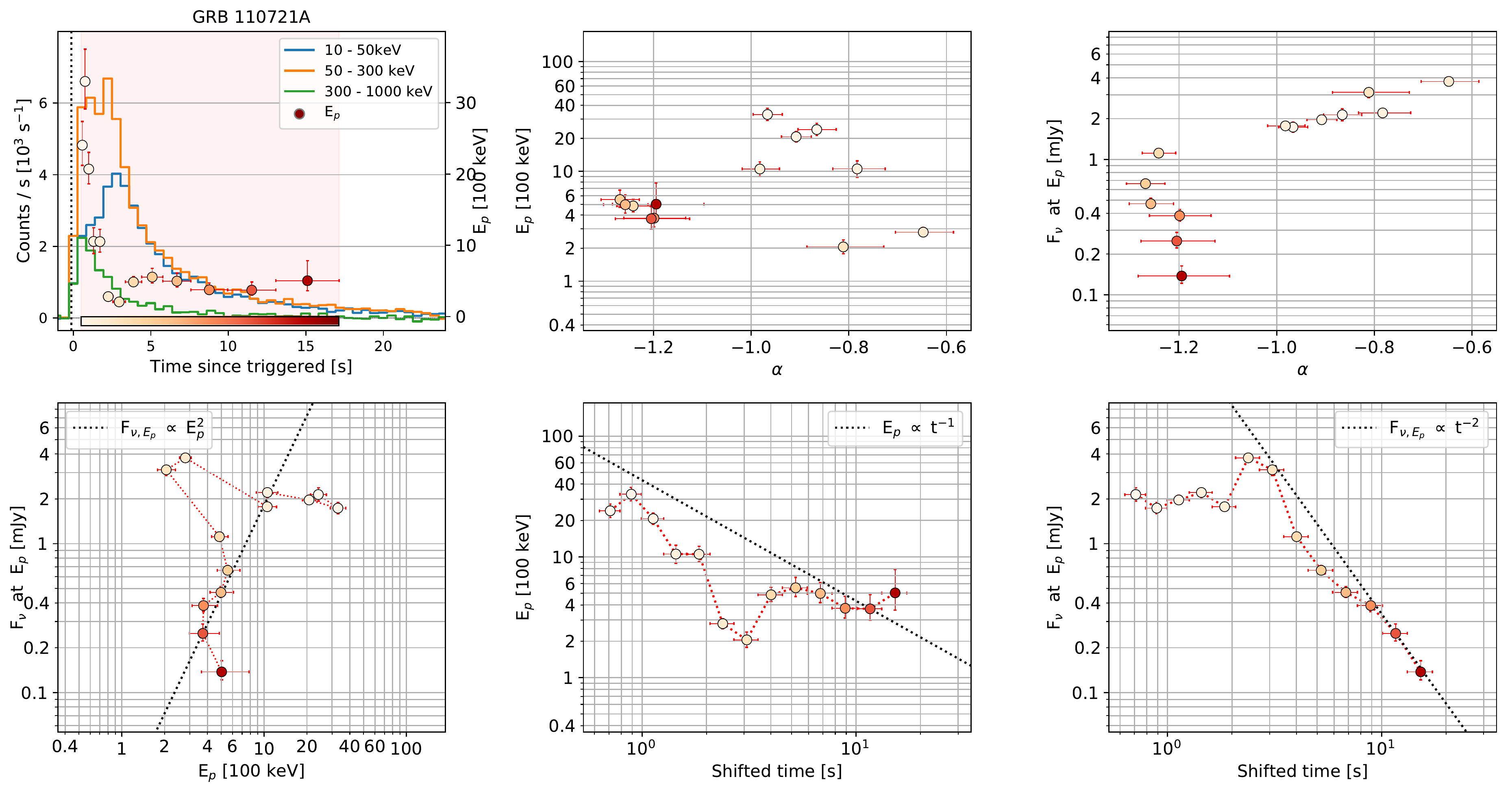} \\ 
\caption{Spectral analysis on GRB\,110721A}
\label{fig:110721A}
\end{figure}
\pagebreak
\begin{figure}
\centering
\includegraphics[scale=0.45]{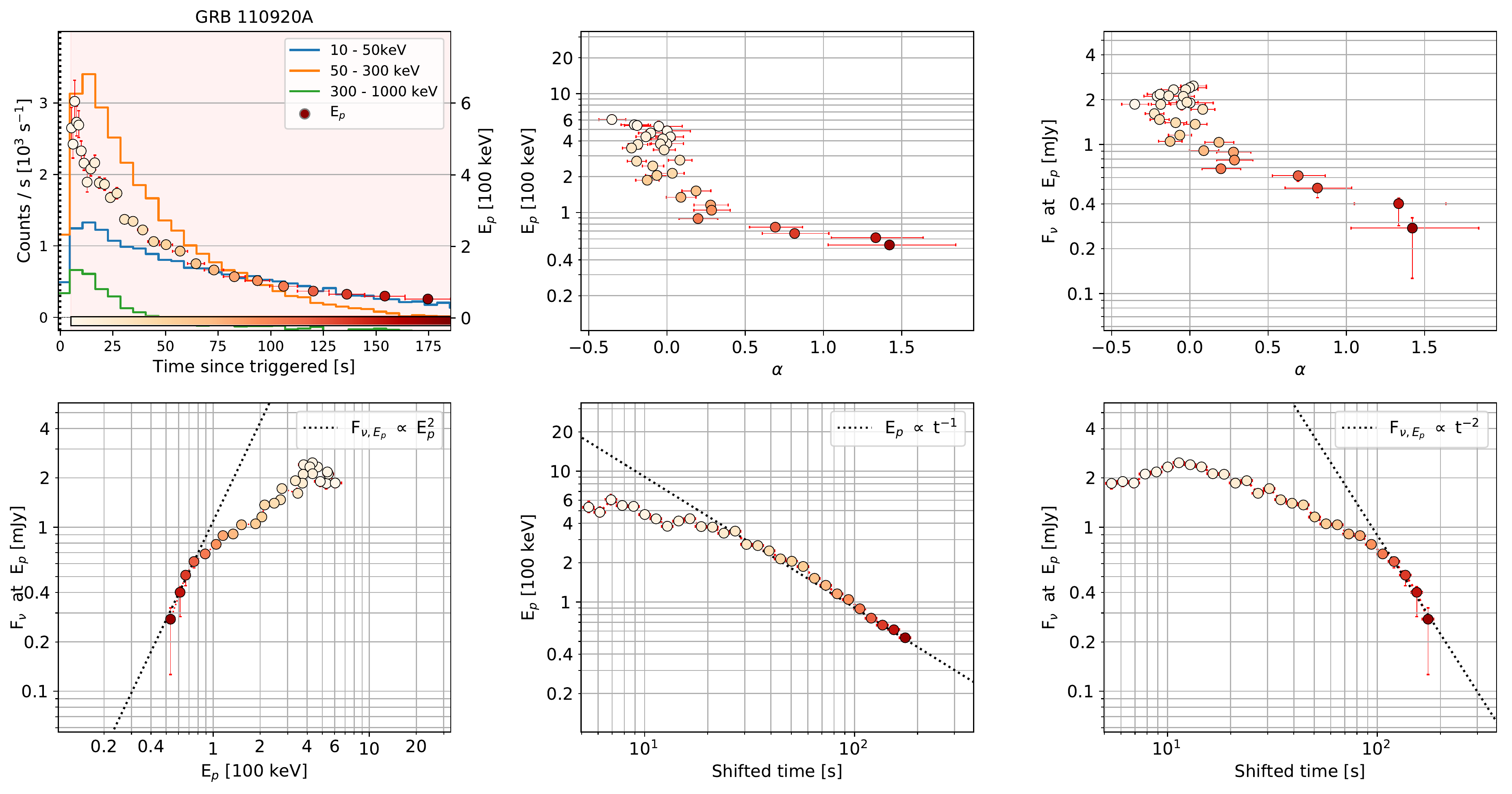} \\ 
\caption{Spectral analysis on GRB\,110920A}
\label{fig:110920A}
\vspace{2cm}
\includegraphics[scale=0.45]{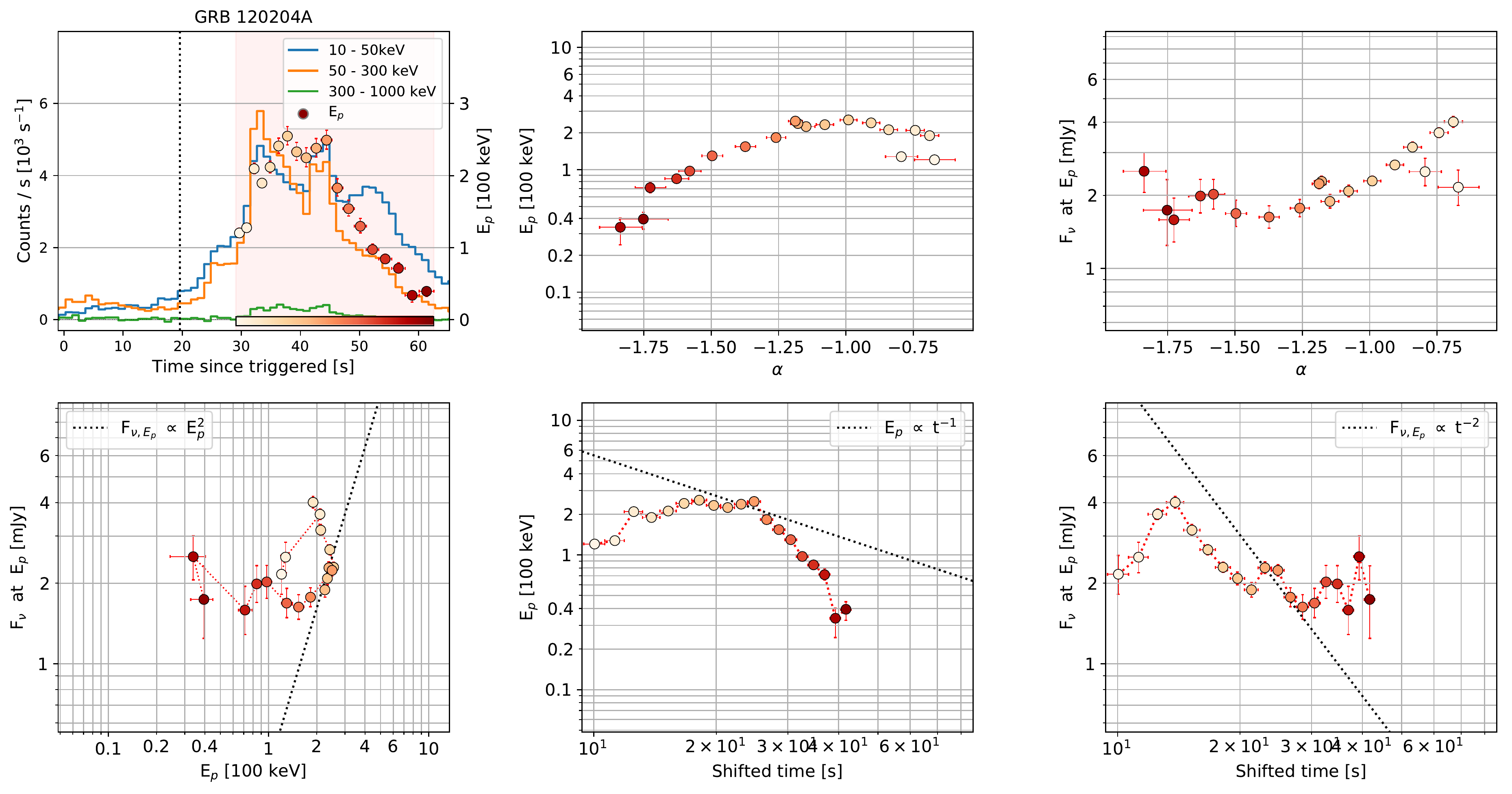} \\ 
\caption{Spectral analysis on GRB\,120204A}
\label{fig:120204A}
\end{figure}
\pagebreak
\begin{figure}
\centering
\includegraphics[scale=0.45]{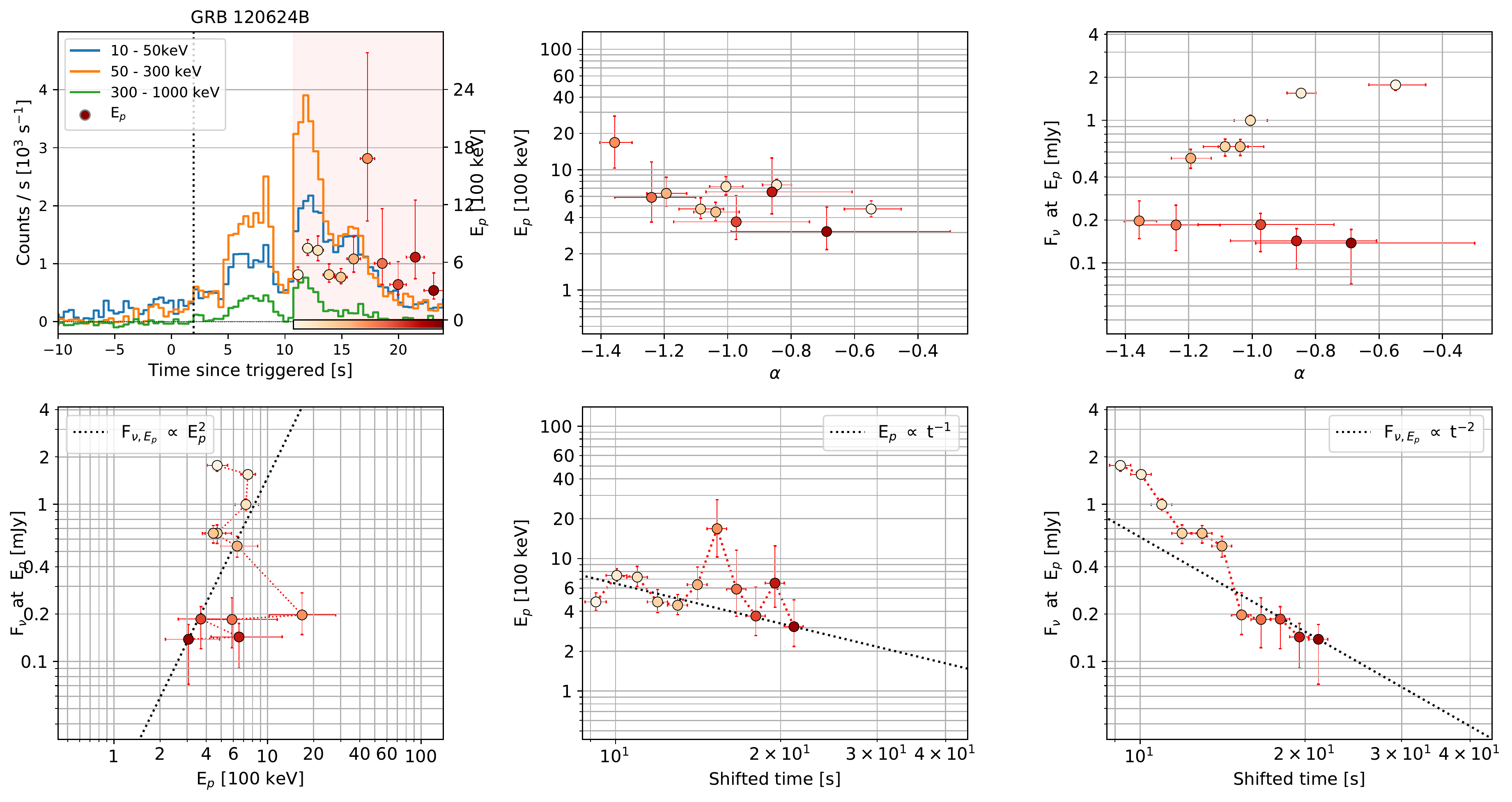} \\ 
\caption{Spectral analysis on GRB\,120624B}
\label{fig:120624B}
\vspace{2cm}
\includegraphics[scale=0.45]{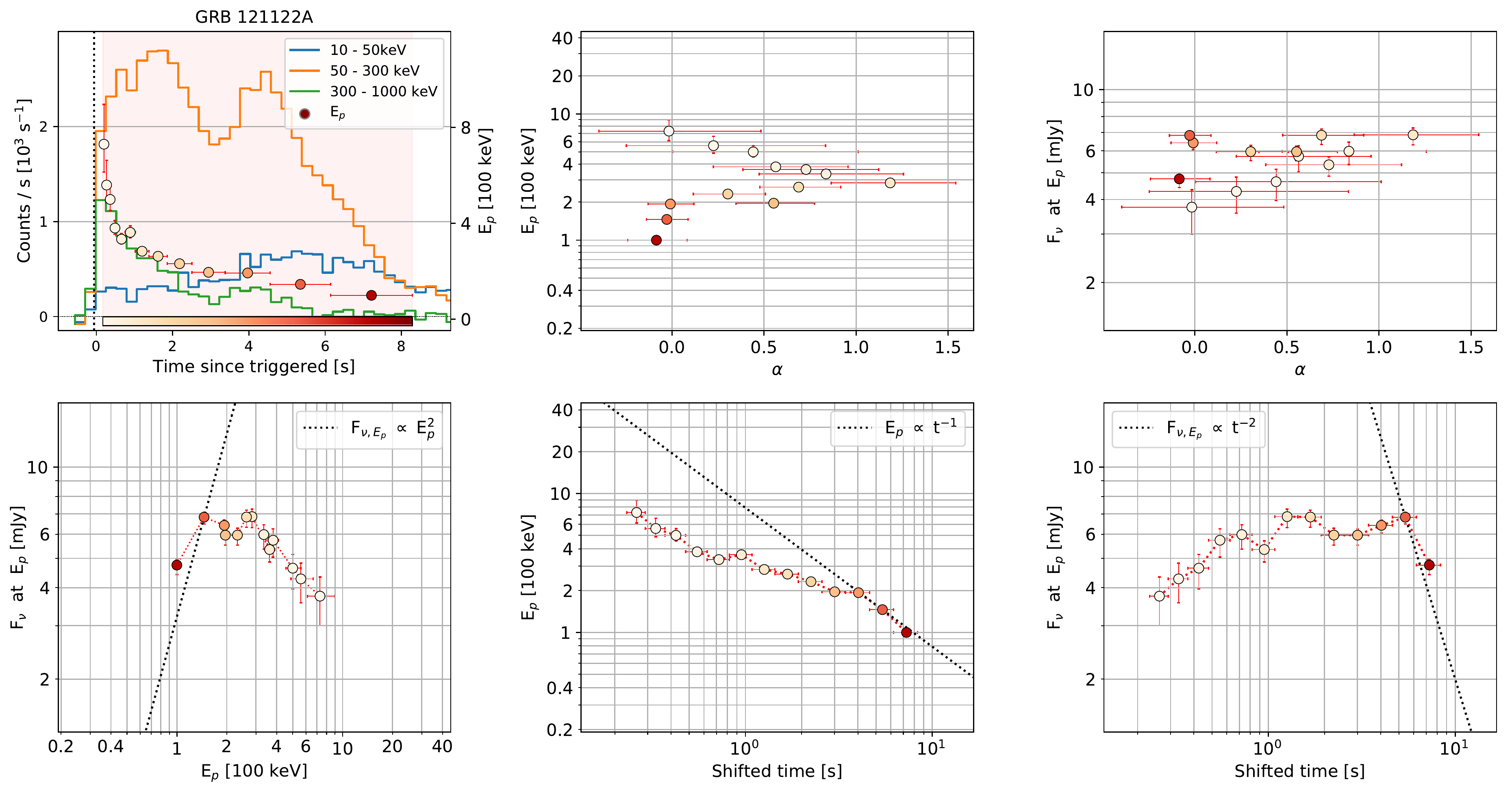} \\ 
\caption{Spectral analysis on GRB\,121122A}
\label{fig:121122A}
\end{figure}
\pagebreak
\begin{figure}
\centering
\includegraphics[scale=0.45]{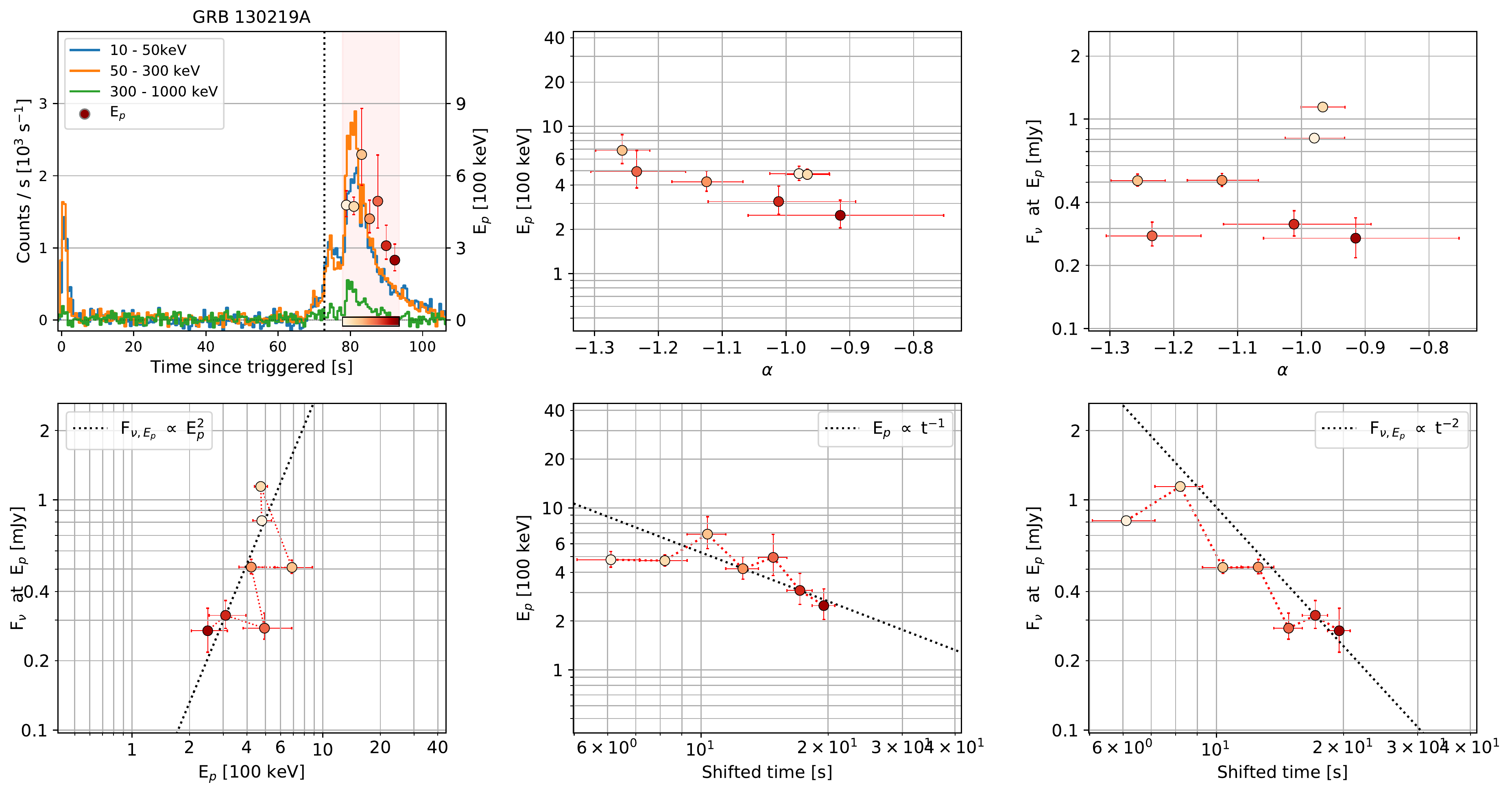} \\ 
\caption{Spectral analysis on GRB\,130219A}
\label{fig:130219A}
\vspace{2cm}
\includegraphics[scale=0.45]{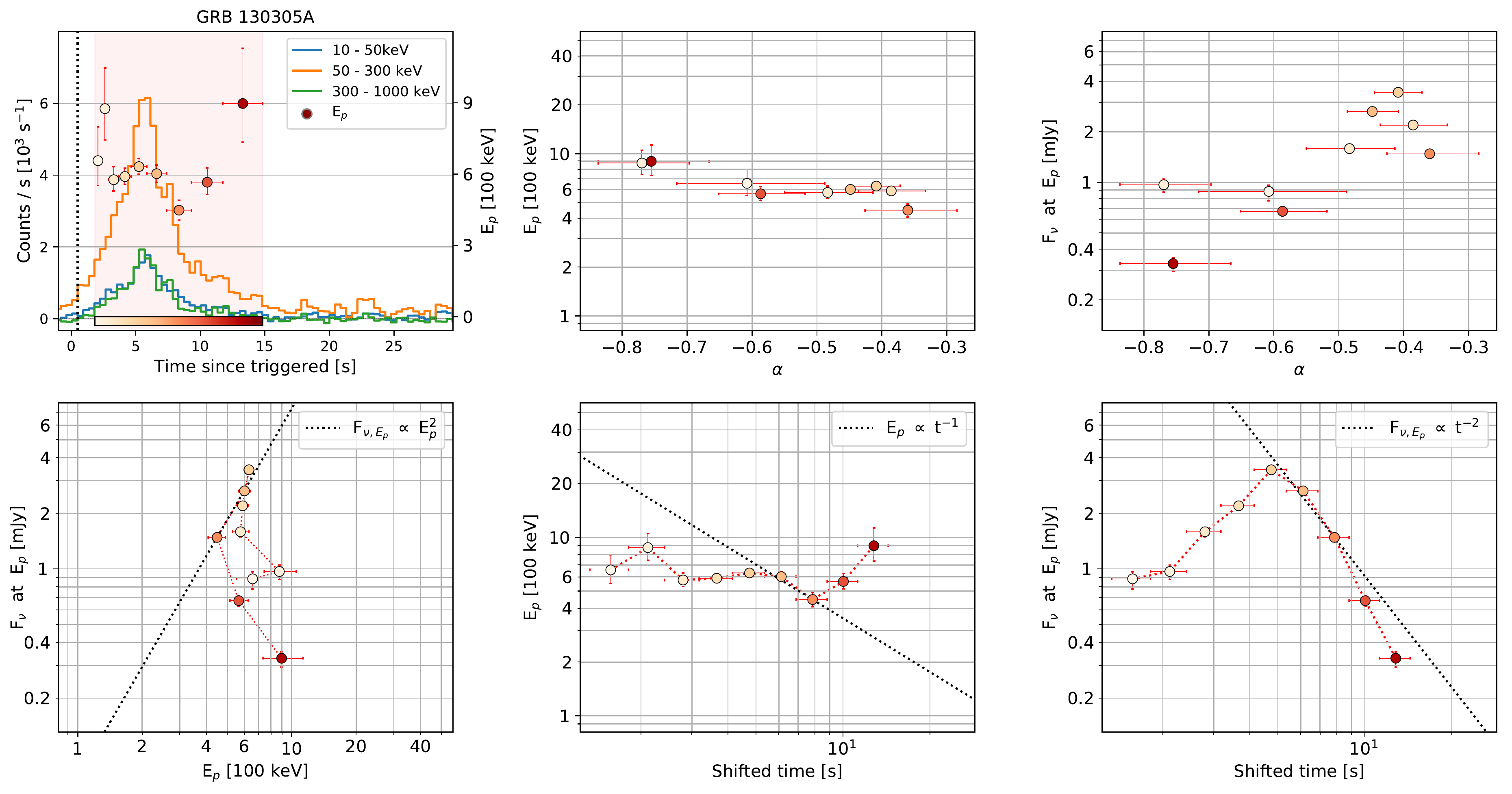} \\ 
\caption{Spectral analysis on GRB\,130305A}
\label{fig:130305A}
\end{figure}
\pagebreak
\begin{figure}
\centering
\includegraphics[scale=0.45]{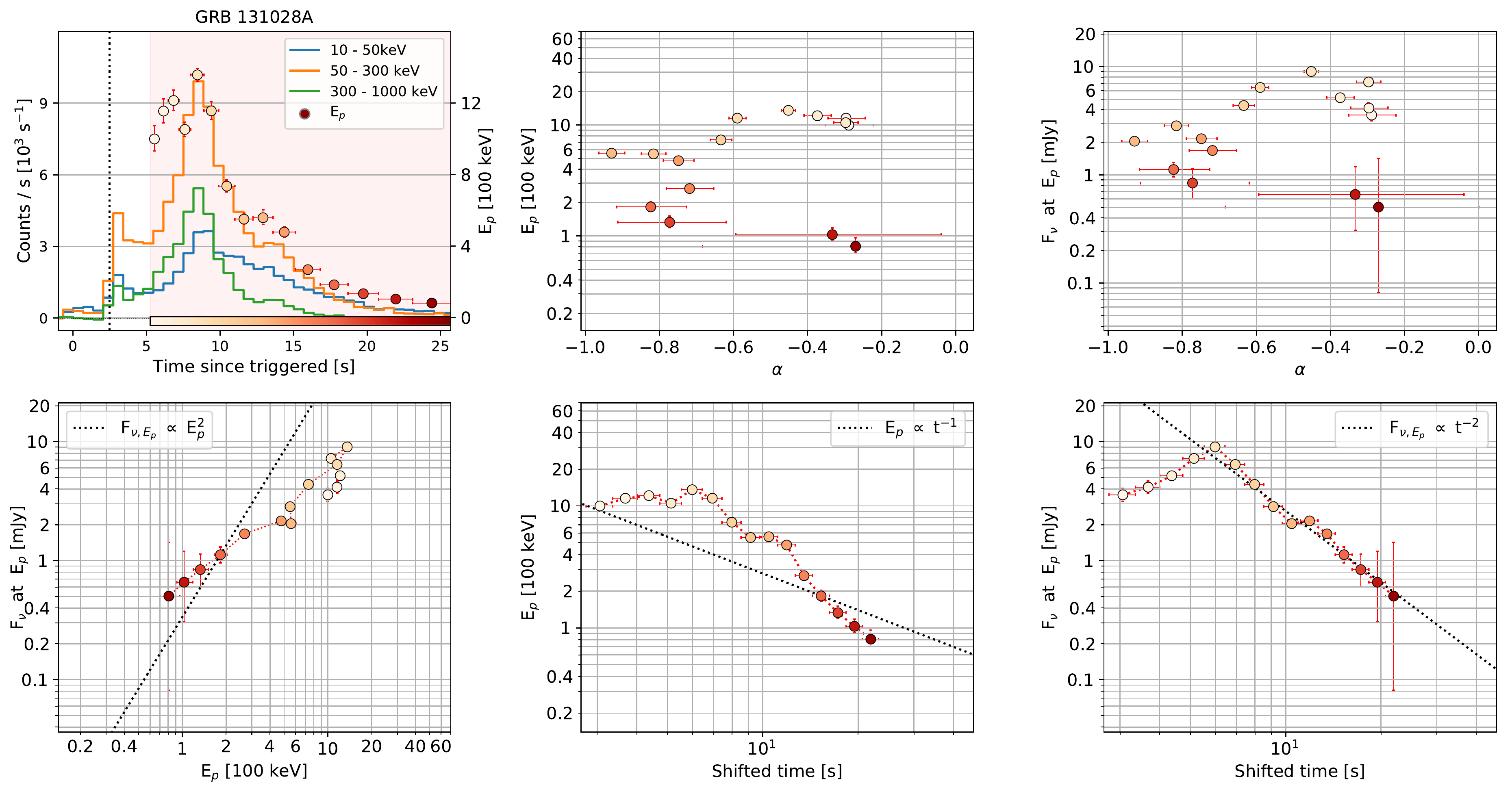} \\ 
\caption{Spectral analysis on GRB\,131028A}
\label{fig:131028A}
\vspace{2cm}
\includegraphics[scale=0.45]{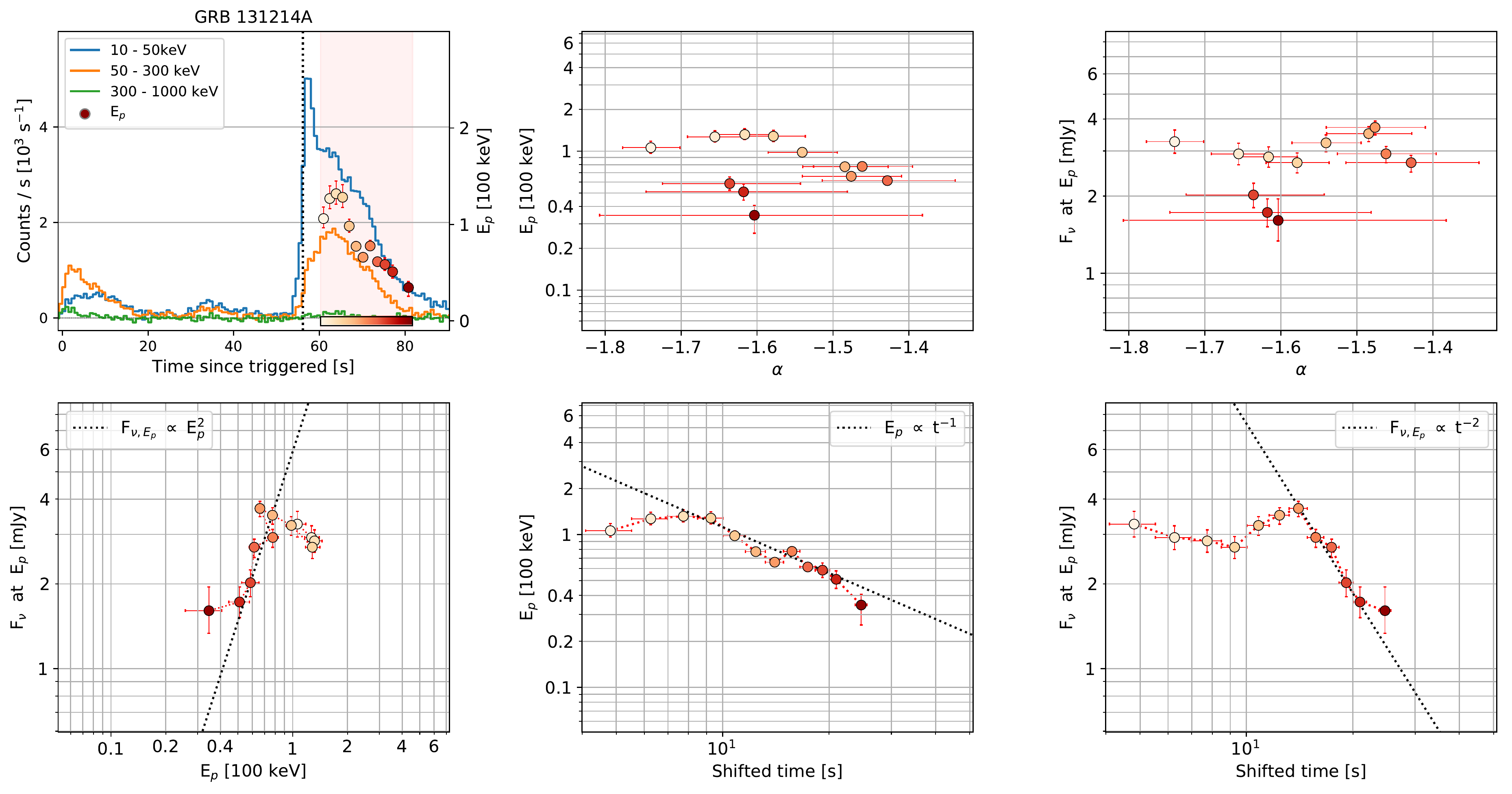} \\ 
\caption{Spectral analysis on GRB\,131214A}
\label{fig:131214A}
\end{figure}
\pagebreak
\begin{figure}
\centering
\includegraphics[scale=0.45]{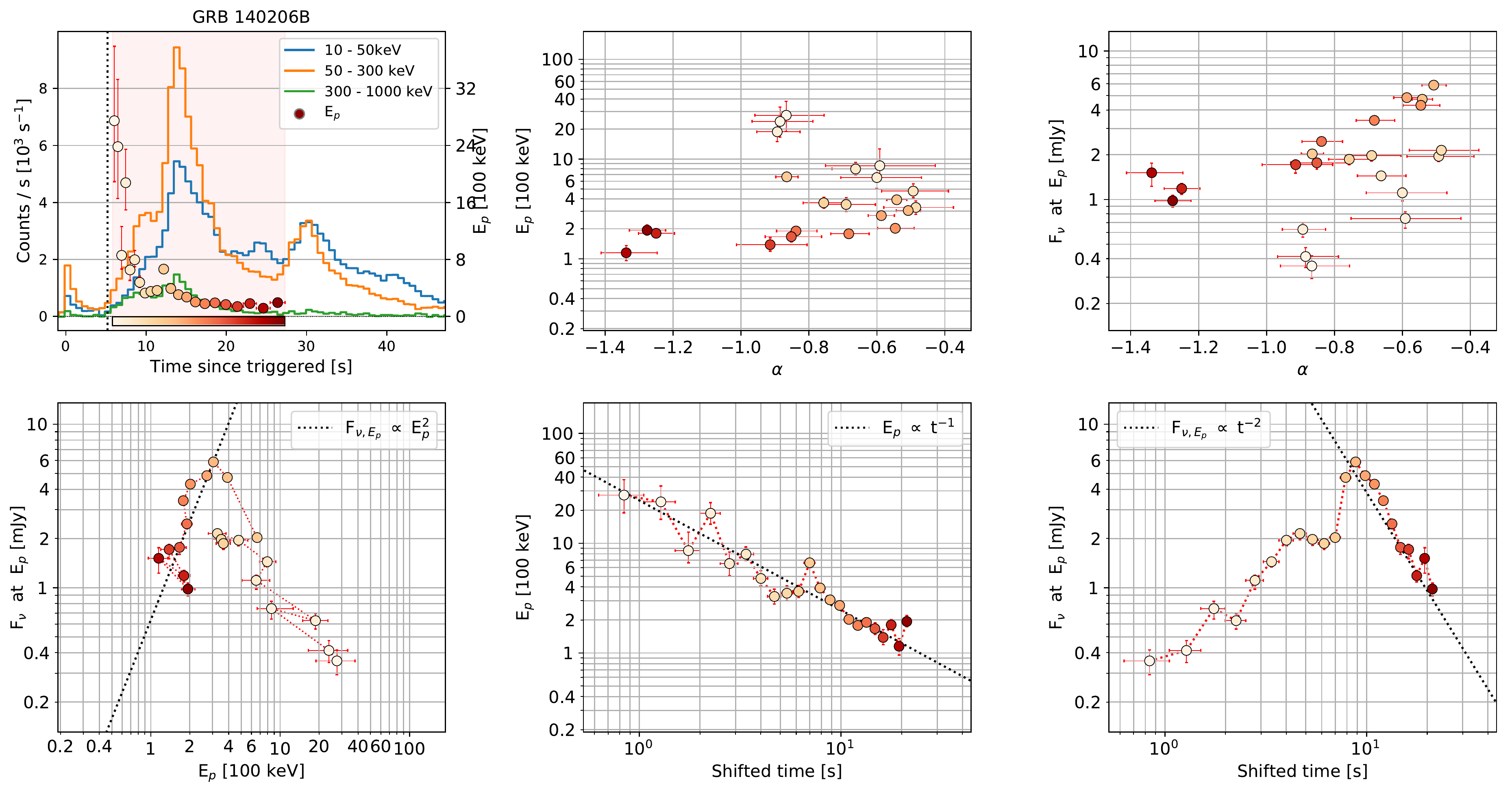} \\ 
\caption{Spectral analysis on GRB\,140206B}
\label{fig:140206B}
\vspace{2cm}
\includegraphics[scale=0.45]{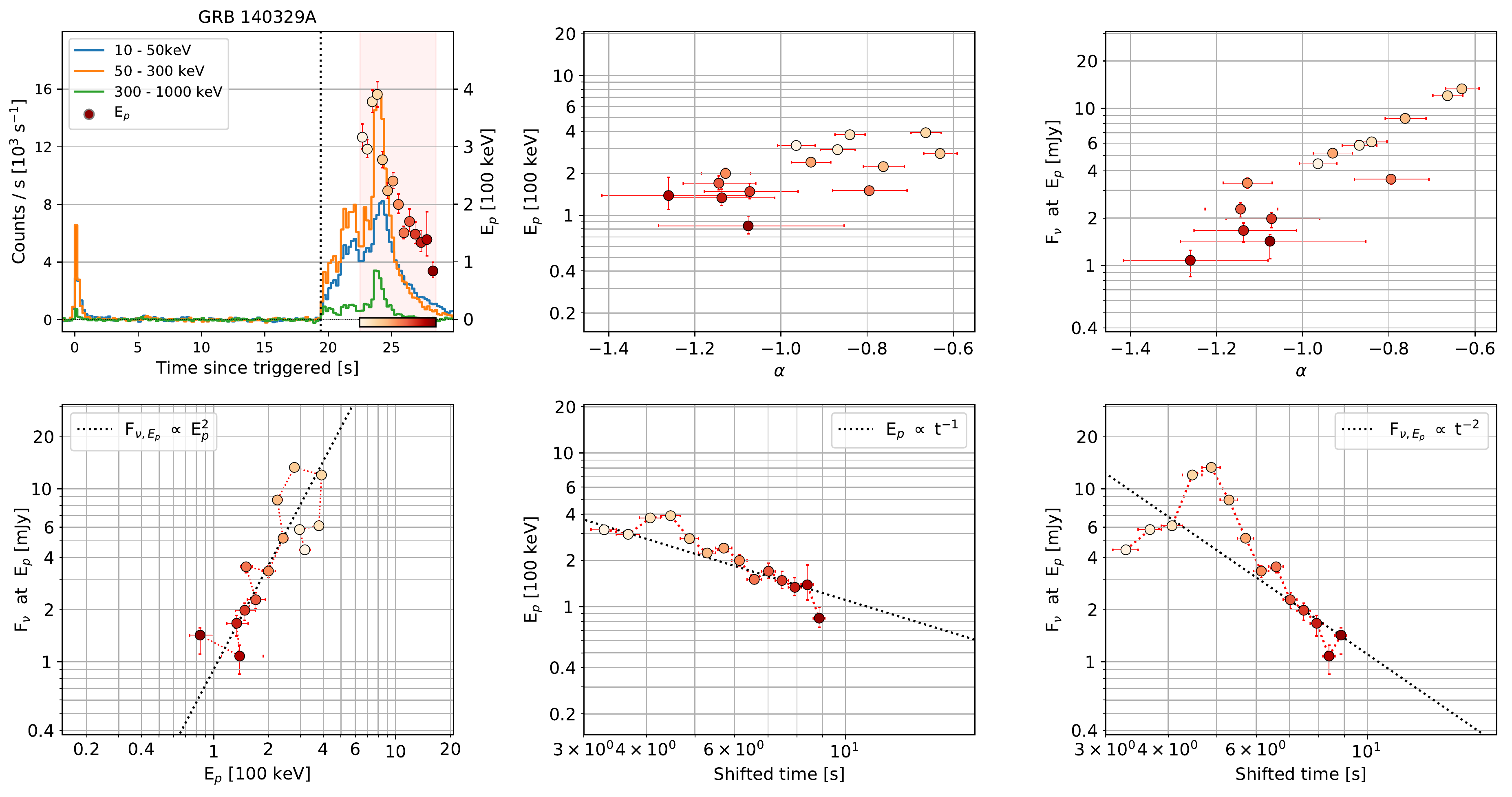} \\ 
\caption{Spectral analysis on GRB\,140329A}
\label{fig:140329A}
\end{figure}
\pagebreak
\begin{figure}
\centering
\includegraphics[scale=0.45]{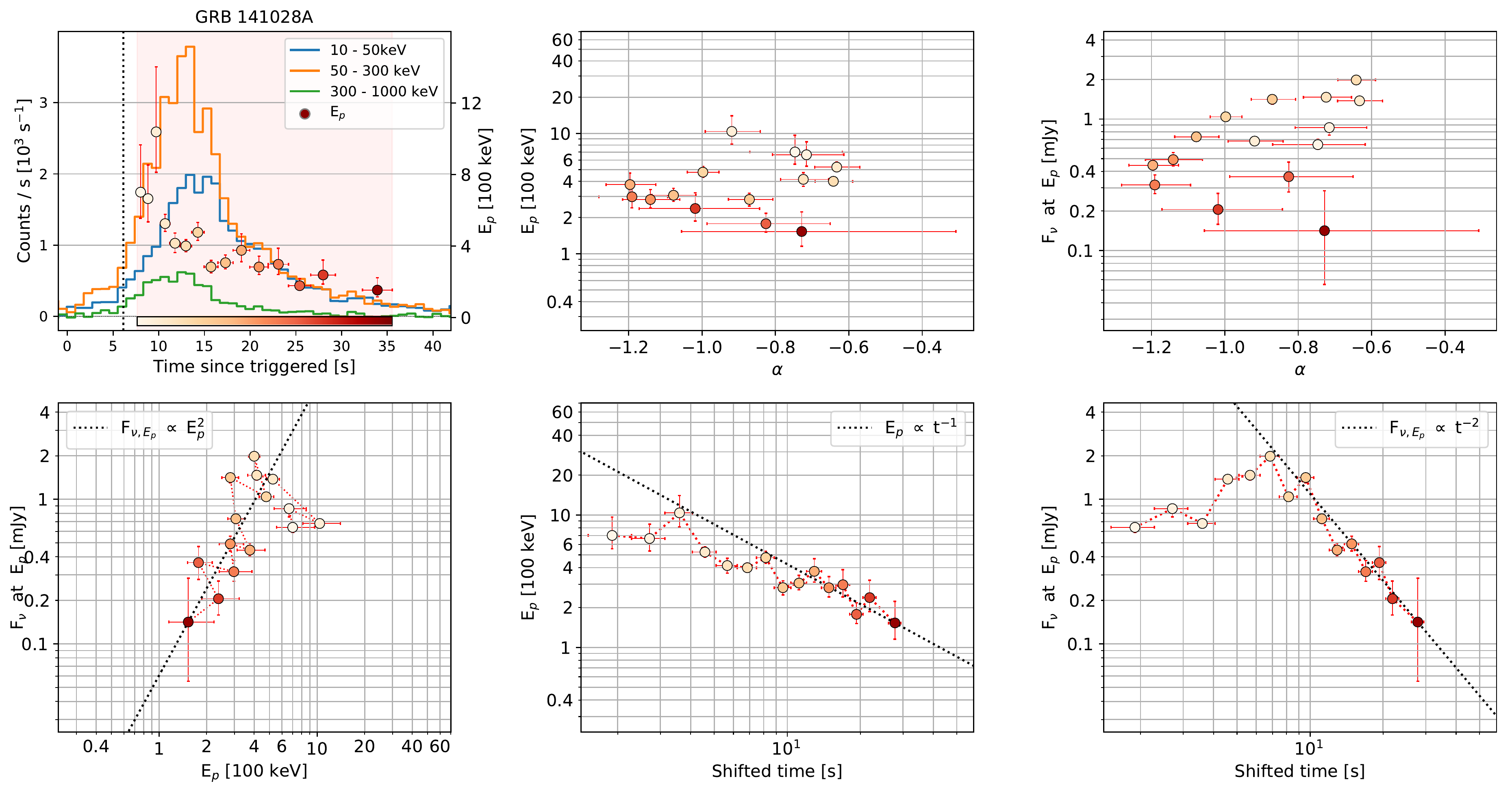} \\ 
\caption{Spectral analysis on GRB\,141028A}
\label{fig:141028A}
\vspace{2cm}
\includegraphics[scale=0.45]{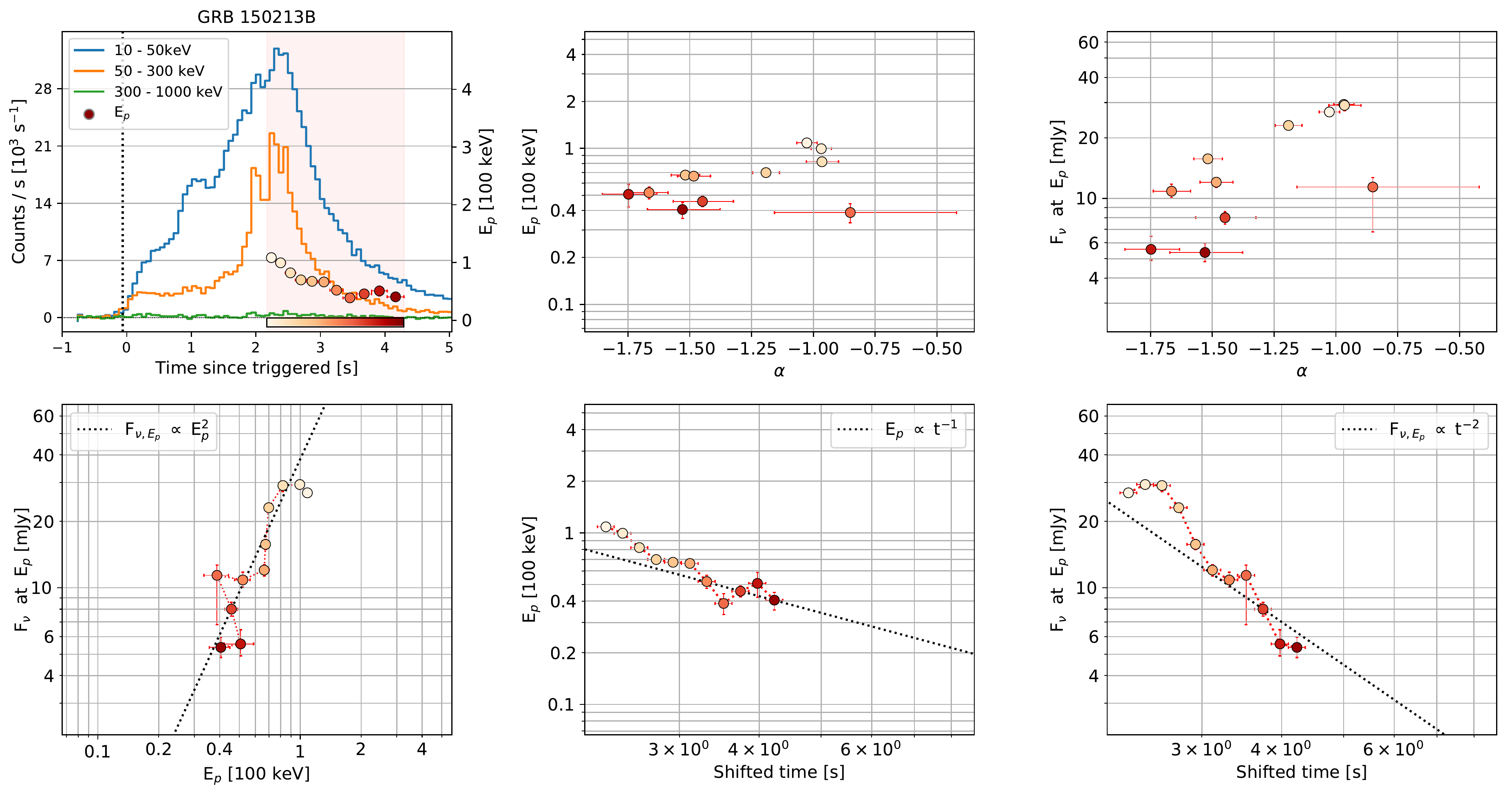} \\ 
\caption{Spectral analysis on GRB\,150213B}
\label{fig:150213B}
\end{figure}
\pagebreak
\begin{figure}
\centering
\includegraphics[scale=0.45]{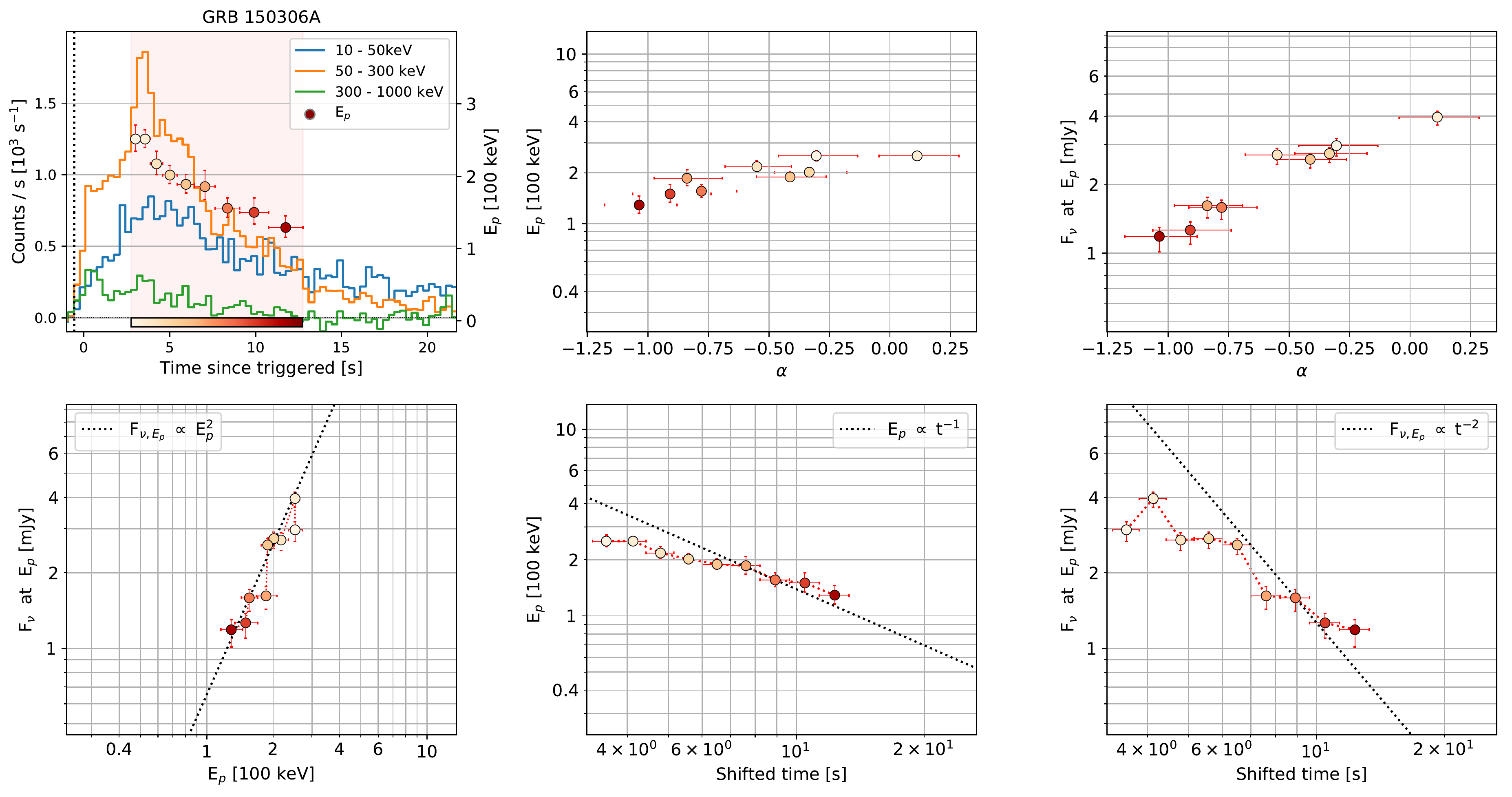} \\ 
\caption{Spectral analysis on GRB\,150306A}
\label{fig:150306A}
\vspace{2cm}
\includegraphics[scale=0.45]{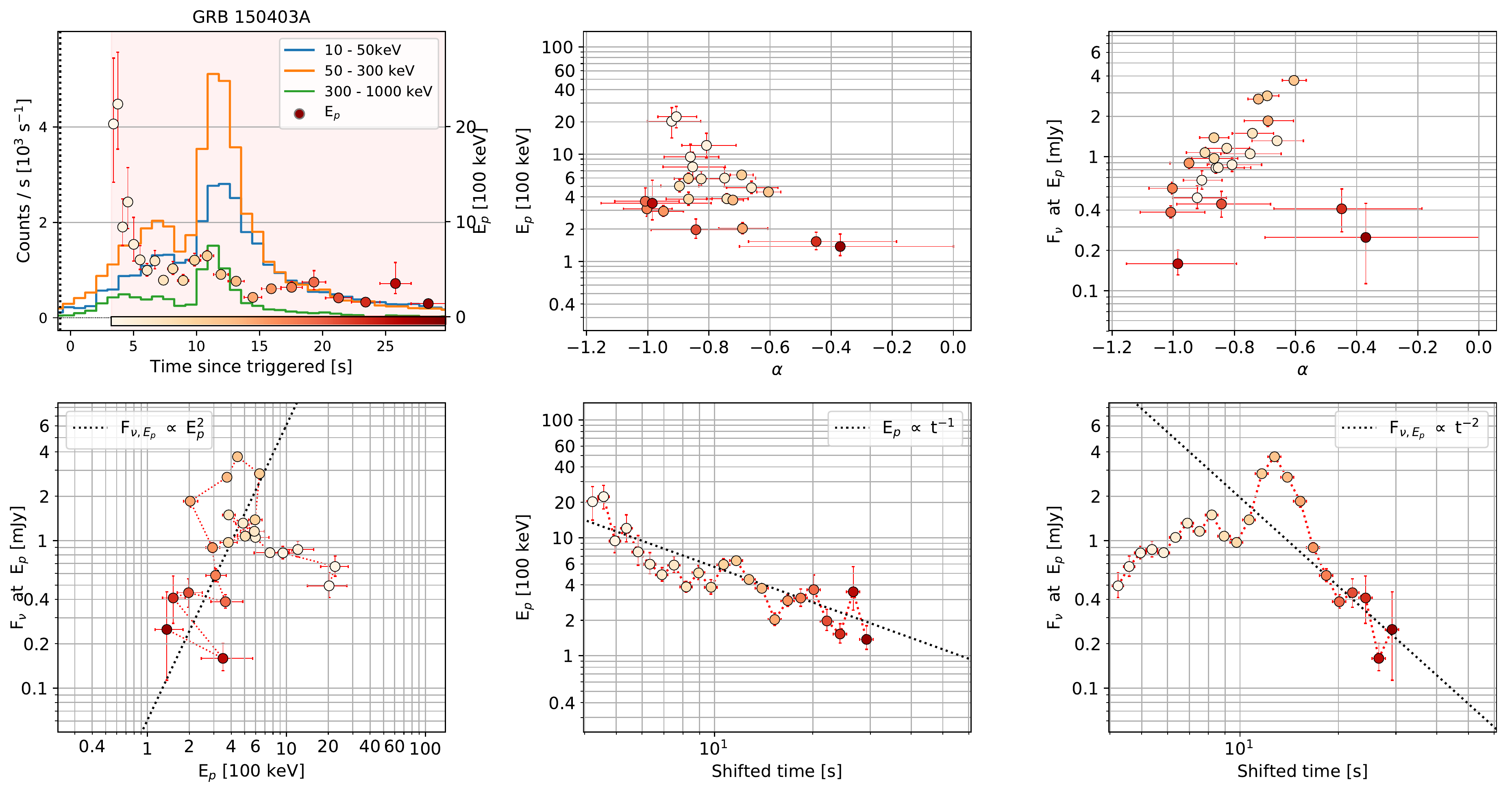} \\ 
\caption{Spectral analysis on GRB\,150403A}
\label{fig:150403A}
\end{figure}
\pagebreak
\begin{figure}
\centering
\includegraphics[scale=0.45]{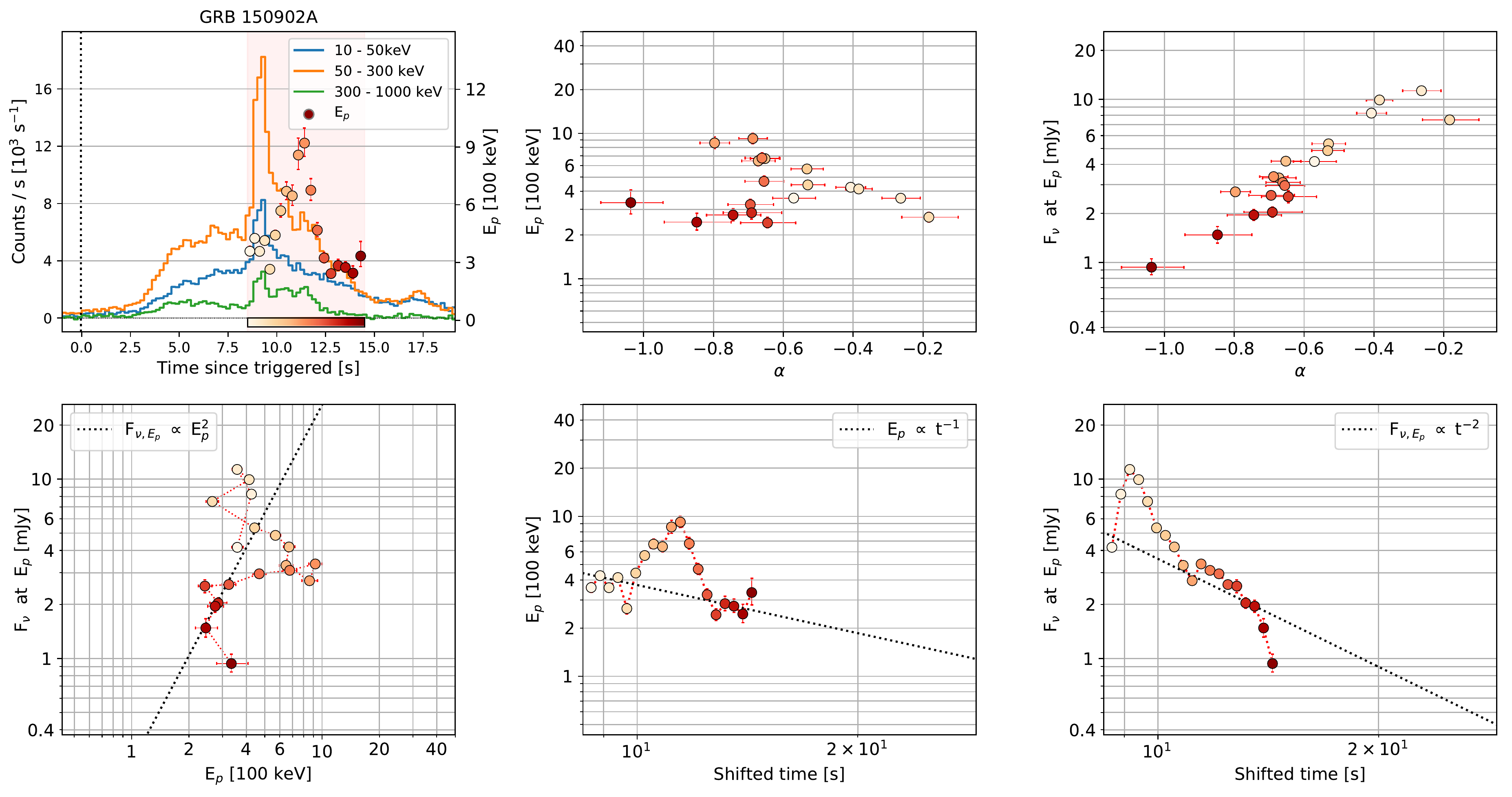} \\ 
\caption{Spectral analysis on GRB\,150902A}
\label{fig:150902A}
\vspace{2cm}
\includegraphics[scale=0.45]{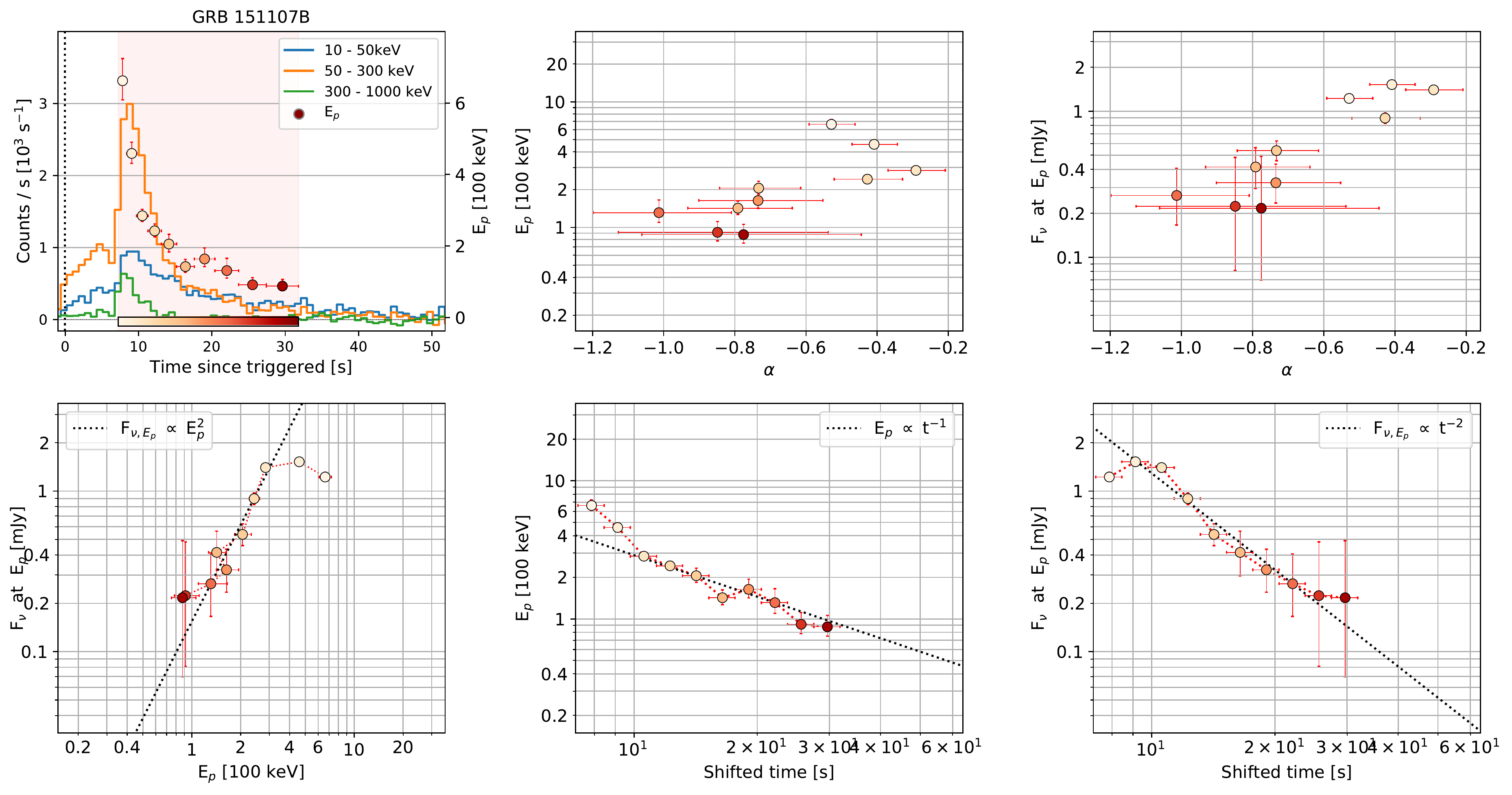} \\ 
\caption{Spectral analysis on GRB\,151107B}
\label{fig:151107B}
\end{figure}
\pagebreak
\begin{figure}
\centering
\includegraphics[scale=0.45]{160113398.pdf} \\ 
\caption{Spectral analysis on GRB\,160113A}
\label{fig:160113A}
\vspace{2cm}
\includegraphics[scale=0.45]{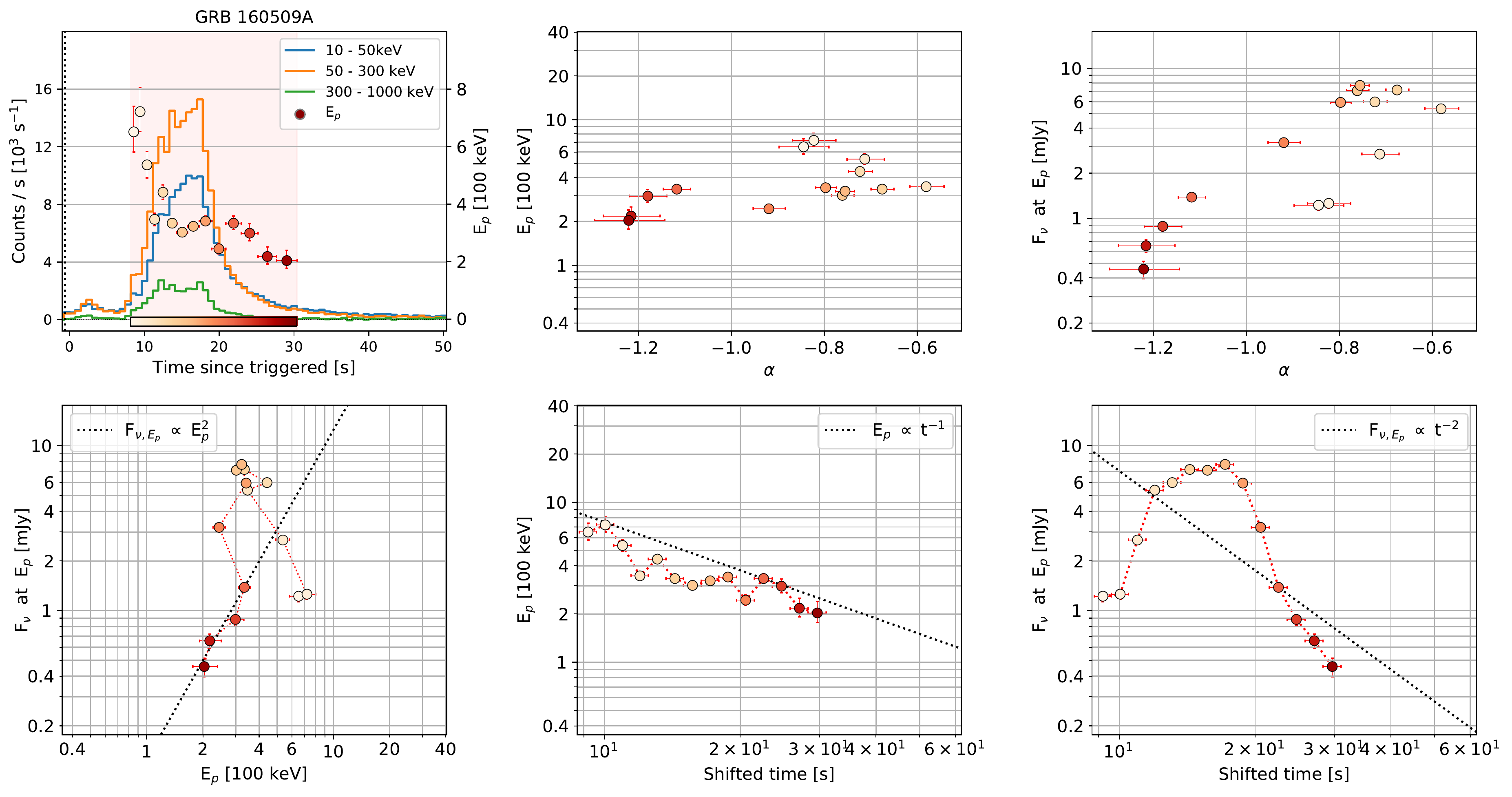} \\ 
\caption{Spectral analysis on GRB\,160509A}
\label{fig:160509A}
\end{figure}
\pagebreak
\begin{figure}
\centering
\includegraphics[scale=0.45]{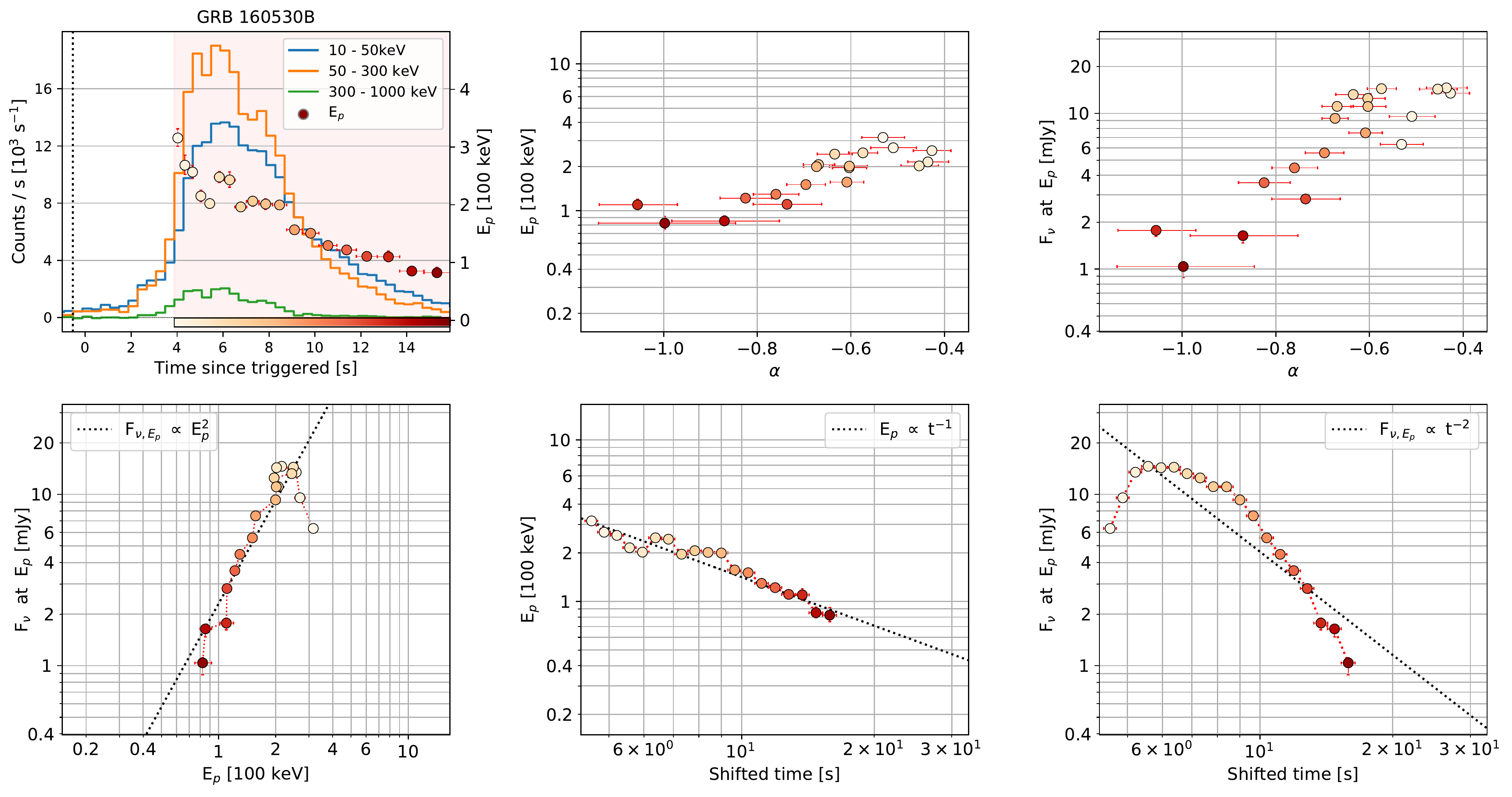} \\ 
\caption{Spectral analysis on GRB\,160530B}
\label{fig:160530B}
\vspace{2cm}
\includegraphics[scale=0.45]{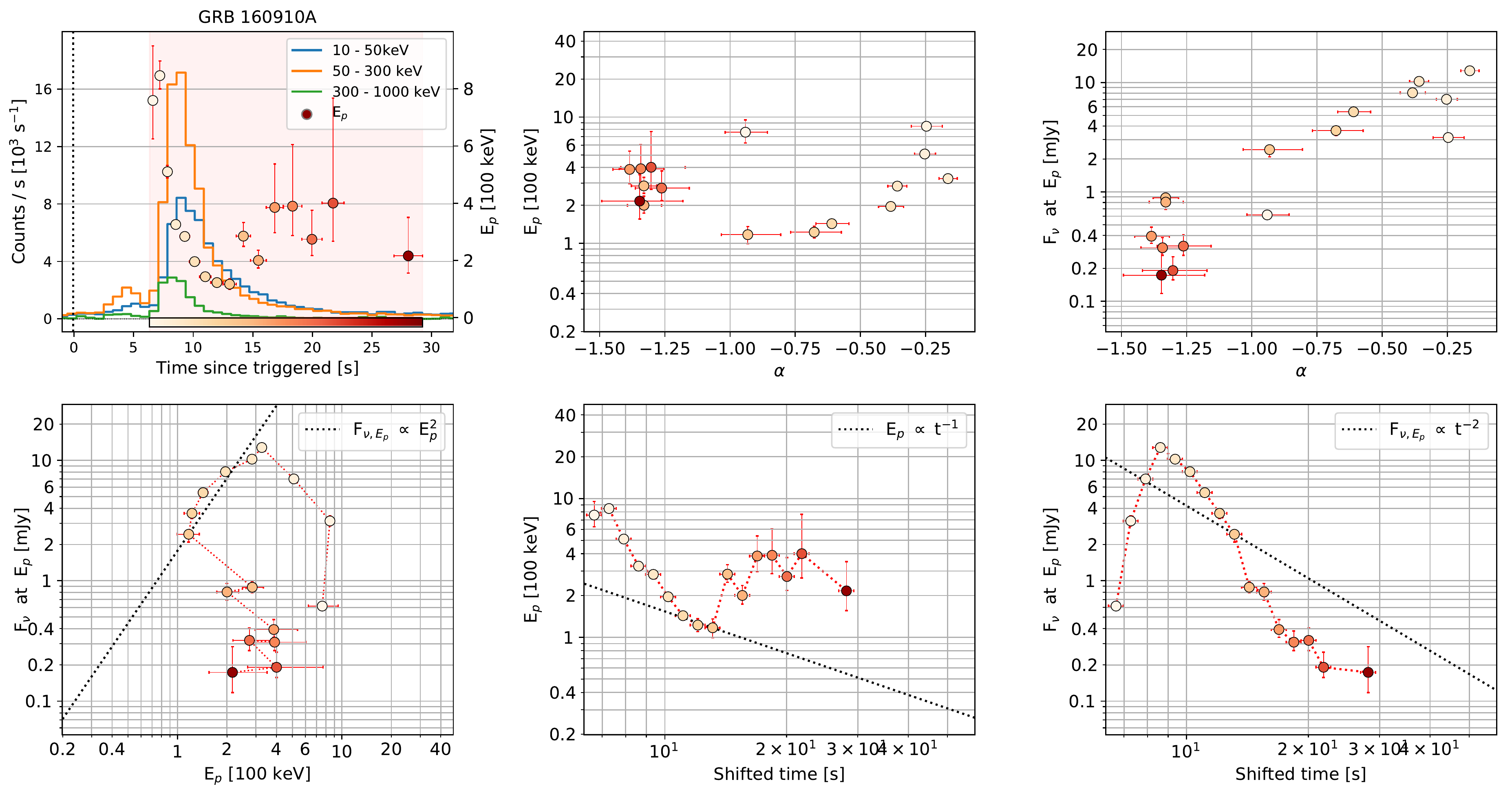} \\ 
\caption{Spectral analysis on GRB\,160910A}
\label{fig:160910A}
\end{figure}
\pagebreak
\begin{figure}
\centering
\includegraphics[scale=0.45]{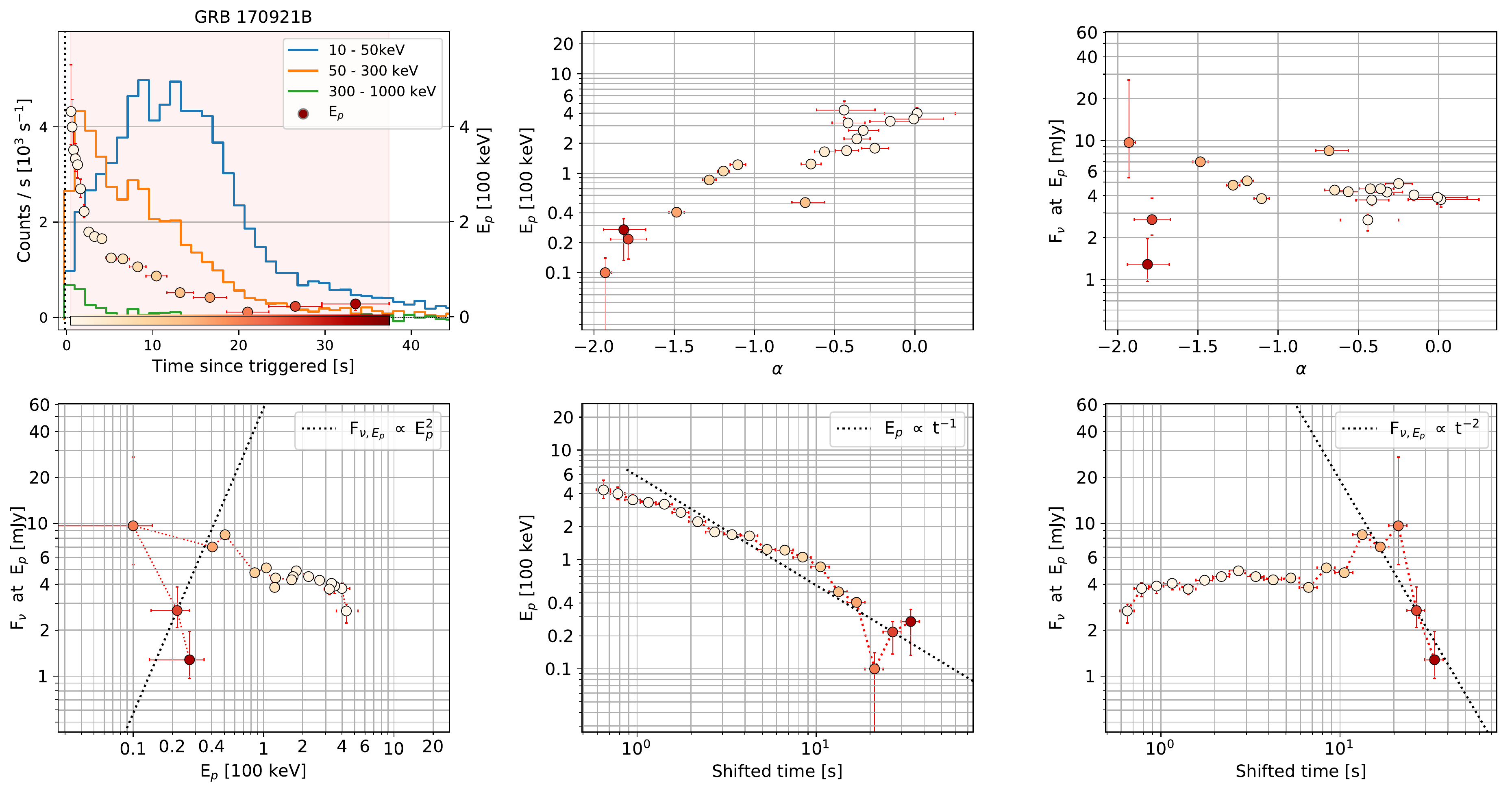} \\ 
\caption{Spectral analysis on GRB\,170921B}
\label{fig:170921B}
\vspace{2cm}
\includegraphics[scale=0.45]{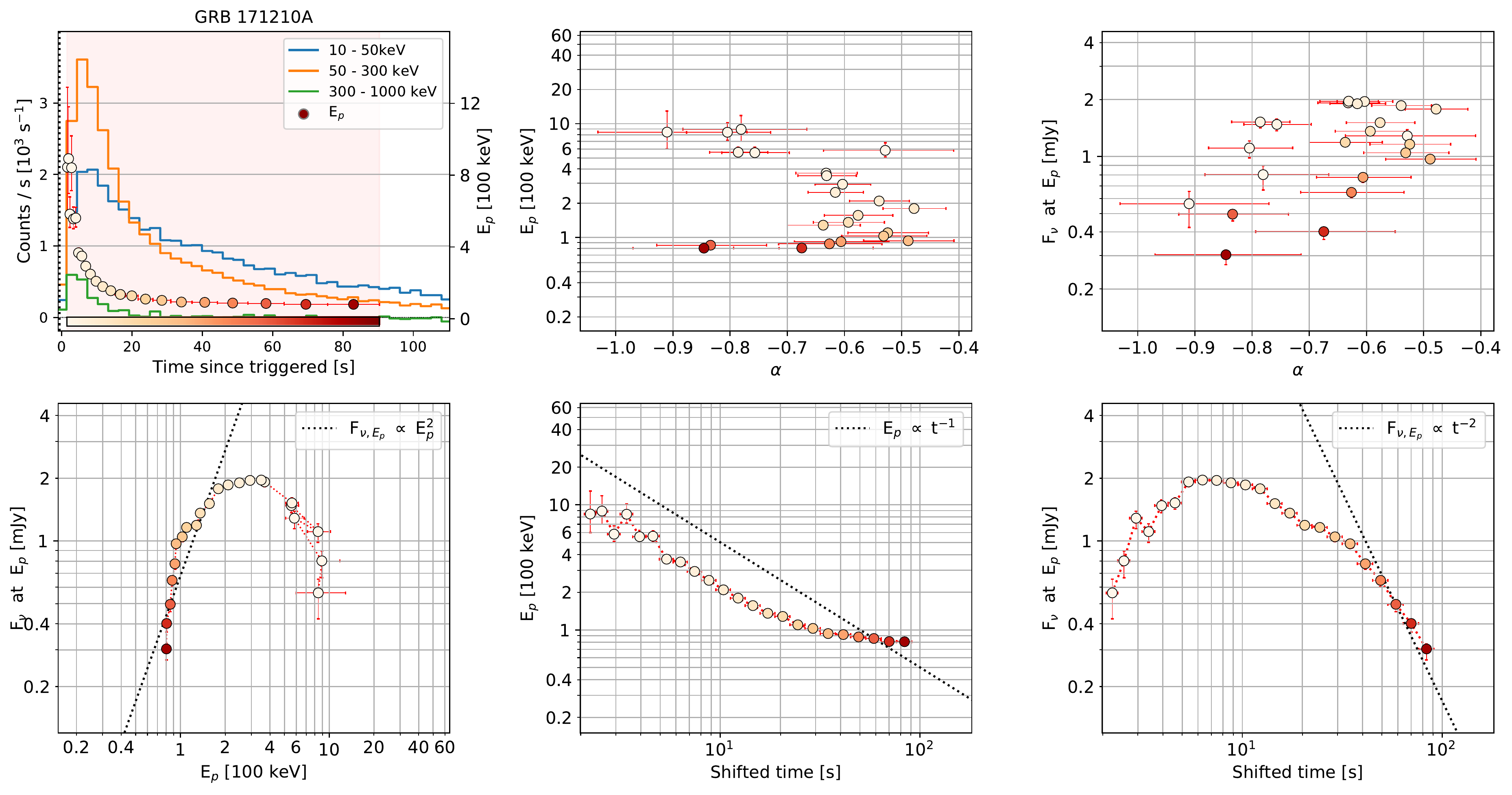} \\ 
\caption{Spectral analysis on GRB\,171210A}
\label{fig:171210A}
\end{figure}
\end{document}